\documentclass[fleqn,usenatbib,useAMS]{mnras}
\usepackage{graphicx}	% Including figure files
\usepackage{amsmath}	% Advanced maths commands
\usepackage{amssymb}	% Extra maths symbols
\usepackage{multicol}        % Multi-column entries in tables
\usepackage{bm}		% Bold maths symbols, including upright Greek
\usepackage{pdflscape}	% Landscape pages

\usepackage[export]{adjustbox}

\usepackage{comment}
\usepackage{xpatch,algpseudocode} %for algorithms
\usepackage{booktabs} % Fancy, professional, tables
\usepackage{tabularx} % Fancy, professional, tables
\usepackage{siunitx}

\usepackage{booktabs,amsfonts,dcolumn}
\newcolumntype{d}[1]{D..{#1}}
\newcolumntype{R}[1]{>{\raggedleft\arraybackslash}p{#1}}
\newcolumntype{L}[1]{>{\raggedright\arraybackslash}p{#1}}
\newcolumntype{C}{>{\centering\arraybackslash}X}

\usepackage{multicol}
\usepackage{lineno}
\usepackage{natbib}
\usepackage[T1]{fontenc}
\usepackage{upquote}
\usepackage{ae,aecompl}
\PassOptionsToPackage{normalem}{ulem}
\usepackage{ulem}

\usepackage{txfonts}

\usepackage{bm}

\usepackage[roman]{parnotes}
\makeatletter
\def\parnoteclear{%
    \gdef\PN@text{}%
    \parnotereset
}
\makeatother

\title[AKFThresh]{Automatic Kalman-Filter-based Wavelet Shrinkage Denoising of 1D Stellar Spectra}

\author[Gilda \& Slepian]{
Sankalp Gilda$^{1}$\thanks{Contact e-mail: \href{mailto:s.gilda@ufl.edu}{s.gilda@ufl.edu}} \&
Zachary Slepian$^{1,2,3}$
\\
$^{1}$Department of Astronomy, University of Florida, 211 Bryant Space Science Center, Gainesville, FL 32611-2055, USA\\
$^{2}$Lawrence Berkeley National Laboratory, 1 Cyclotron Road, Berkeley, CA 94720, USA\\
$^{3}$Berkeley Center for Cosmological Physics, University of California, Berkeley, Berkeley, CA 94720, USA
}

\date{Last updated 2019 March 7; in original form 2019 March 7}

\pubyear{2019}

\begin{document}
\label{firstpage}
\pagerange{\pageref{firstpage}--\pageref{lastpage}}
\maketitle

\begin{abstract}
We propose a non-parametric method to denoise 1D stellar spectra based on wavelet shrinkage followed by adaptive Kalman thresholding. Wavelet shrinkage denoising involves applying the Discrete Wavelet Transform (DWT) to the input signal, `shrinking' certain frequency components in the transform domain, and then applying inverse DWT to the reduced components. The performance of this procedure is influenced by the choice of base wavelet, the number of decomposition levels, and the thresholding function. Typically, these parameters are chosen by `trial and error', which can be strongly dependent on the properties of the data being denoised. We here introduce an adaptive Kalman-filter-based thresholding method that eliminates the need for choosing the number of decomposition levels. We use the `Haar' wavelet basis, which we found to provide excellent filtering for 1D stellar spectra, at a low computational cost. We introduce various levels of Poisson noise into synthetic PHOENIX spectra, and test the performance of several common denoising methods against our own. It proves superior in terms of noise suppression and peak shape preservation. We expect it may also be of use in automatically and accurately filtering low signal-to-noise galaxy and quasar spectra obtained from surveys such as SDSS, Gaia, LSST, PESSTO, VANDELS, LEGA-C, and DESI.

%This is a guide for preparing papers for \textit{Monthly Notices of the Royal Astronomical Society} using the \verb'mnras' \LaTeX\ package.
%It provides instructions for using the additional features in the document class.
%This is not a general guide on how to use \LaTeX, and nor does it replace the journal's instructions to authors.
%See \texttt{mnras\_template.tex} for a simple template.
\end{abstract}


%%%%%%%%%%%%%%%%%%%%%%%%%%%%%%%%%%%%%%%%%%%%%%%%%%

%%%%%%%%%%%%%%%%% BODY OF PAPER %%%%%%%%%%%%%%%%%%

% The MNRAS class isn't designed to include a table of contents, but for this document one is useful.
% I therefore have to do some kludging to make it work without masses of blank space.
%\begingroup
%\let\clearpage\relax
%\tableofcontents
%\endgroup
%\newpage
%\linenumbers
\section{Introduction}
Most commonly-used methods for spectral denoising are parametric and thus require laborious user intervention for selection of appropriate parameters. The most popular among them is the Fourier transform (FT). Unlike other time-based techniques for noise removal such as the moving average (MA) and the exponential moving average (EMA), the FT provides a vision of a signal in the complementary domain (i.e., frequency domain, if the original signal is a time series). Because of the periodicity of the sinusoids used as the basis functions in the FT, it is well-suited to processing stationary signals. However, it has several drawbacks. It is inefficient when dealing with a non-stationary signal with varying shape and width, has zero resolution in the original (time) domain, and requires a user-specified cut-off frequency to estimate the noise in the signal. While the time-windowed FT offers an improvement by introducing a moving window that represents a compromise between time and frequency resolution, it is still limited by use of a fixed window size \citep{kaiser2010friendly}. 

Wavelet transform (WT), on the other hand, is able to achieve an excellent trade-off between resolutions in time and frequency domains, by providing high (low) resolution in the frequency (time) domain for small frequency values, and high (low) resolution in the time (frequency) domain for large frequency values. This provides a clear advantage over FT and short time-windowed FT for exploring features of interest in the input signal in both time and frequency domains. However, traditional denoising approaches built using wavelet transform still require trial-and-error by the user to determine and suppress the noise in the signal, as we discuss further in \S\ref{wavelet} and \S\ref{limitations}. 

In this work, we present a new denoising approach, based on wavelet decomposition \citep{graps1995introduction} and adaptive Kalman-filter-based thresholding of the resultant components, which significantly improves denoising performance. It also provides a simpler, quicker implementation relative to previous methods, including FT, by being effectively non-parametric and hence independent of user input. For characterization of an object (e.g. a star, galaxy, or quasar) based on its spectrum, the precision and accuracy of the derived parameters depend strongly on two factors---the spectral resolution, and the signal-to-noise ratio (SNR). While the former is completely dependent on hardware (spectrograph), the latter can be controlled to some extent with software (algorithms). This dependence of object characterization on spectral SNR becomes even more prominent when dealing with low SNR spectra, as is common in large galaxy surveys. Our proposed method recursively decomposes the input stellar absorption spectrum---corrupted with Poisson (shot) noise---into high-frequency and low-frequency signals via a method called the wavelet transform. These high-frequency signals are then denoised (i.e. filtered) using Kalman filters. Kalman filters use a heuristic to derive noise statistics, and by employing a predictor-corrector scheme iteratively filter the signals. The denoised high-frequency components are then re-combined with the low-frequency signal at the last level, and an inverse wavelet transform used to recover the denoised stellar spectrum. We demonstrate that this method outperforms the conventional signal filtering methods when optimal method parameters are unavailable for the latter. This is indeed likely to be the case in practice due to an absence of a realistic `reference' signal and noise to employ for obtaining optimal parameters for the conventional, parametric methods.

This paper is organized as follows.% In \S\ref{intro_main} we briefly introduce the denoising algorithm proposed in this work, and the statistical tools it uses, namely, the discrete wavelet transform and the Kalman filter. 
 In \S\ref{wavelet} we discuss the wavelet transform and the wavelet shrinkage method. In \S\ref{kalman} we briefly describe the Kalman filter. \S\ref{limitations} describes our choice of wavelets and selection of decomposition levels to denoise, and demonstrates the effectiveness of our proposed method in addressing these issues. In \S\ref{main}, we introduce the synthetic PHOENIX spectra \citep{husser2013new} used for comparing various filtering methods, and provide an in--depth explanation of our proposed algorithm. In \S\ref{results} we use the PHOENIX data to compare the performance of our proposed method with commonly-used filtering methods from the literature. Finally, in \S\ref{conclusions} we comment on our findings and briefly discuss future extensions to the work.

\section{Wavelet-based Signal Denoising}\label{wavelet}

\subsection{Discrete Wavelet Transform}\label{dwt}
The wavelet transform is similar to the Fourier transform (and more so to the windowed Fourier transform) in that both transform the input data into a different domain. However, the main difference between them is this---the Fourier transform decomposes the signal into sines and cosines, i.e. functions localized in Fourier space; in contrast the wavelet transform uses functions (\emph{wavelets}) that are somewhat localized in both real and Fourier space. The sine and cosine functions used in the Fourier transform are `infinite'---in that they never go to zero in the time domain---and hence the signal is deconstructed into waves that are infinitely long. This introduces the `resolution problem'---if we have a high resolution in the frequency domain (i.e. focusing on one frequency in the signal) it is hard to isolate it in time, as each frequency exists across all time. Being uncertain of the time when focusing on the frequency is the flip-side of being uncertain of the frequency when focusing on time.

To overcome this resolution problem a wavelet transform is used to deconstruct the signal. The time-limited quality of wavelets is useful as it provides more resolution in the time domain. Instead of modeling with a wave that has infinite support, one uses a compactly-supported wave convolved with the input signal. To handle different frequencies, the wavelet transform employs varying \emph{scales} for the wavelet. Wavelets in the WT are akin to sinusoidal waves in the FT: just as a sinusoid gets compressed (stretched) for high (low) frequencies, so do wavelets. For the WT, the signal is deconstructed using the same wavelet with different scales, rather than the same sinusoid at different frequencies. As there are hundreds of different wavelets, hundreds of different transforms are possible, each with an associated domain. However each domain has `scale' on the horizontal axis rather than `frequency', and provides better resolution in the time domain by being finite.

In this work, we use a specific type of wavelet transform---the Discrete Wavelet Transform (DWT) (\citealt{bultheel1995learning} gives an in-depth review of wavelet transforms and their properties). DWT is an implementation of the wavelet transform using a discrete set of wavelet scales and translations that obey several rules. DWT is particularly suitable for signal compression and denoising. There are several implementations of the DWT algorithm. The oldest and best-known one---and the one used in this work---is the \citet{mallat1989theory} (pyramidal) algorithm. With this algorithm, the data length is restricted to a power of $2$, requiring padding of the signal vector before applying the DWT. The noisy signal vector is simultaneously passed through a low-pass filter to obtain large scales (analogous to low frequencies in an FT), and a high-pass filter to obtain small scales (analogous to high frequencies in an FT). From the low-pass filter, we get a vector of approximation coefficients $\textbf{A0}$ that represents an estimation of the original signal with half resolution. From the high-pass filter, we obtain a vector of detail coefficients $\textbf{D0}$ that contains the details of the signal. The vector $\textbf{A0}$ can be further decomposed to form a new vector of approximation coefficients $\textbf{A1}$ and a new vector of detail coefficients $\textbf{D1}$.

\begin{figure}
\begin{center}
\includegraphics[width=1\columnwidth]{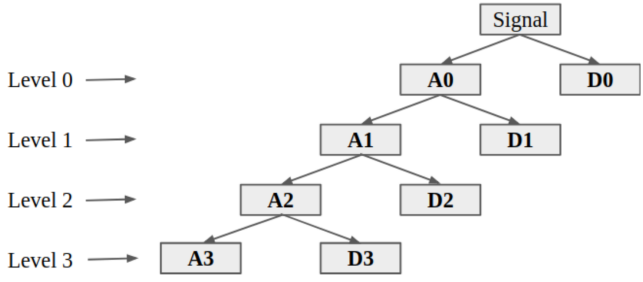}
\caption{Block diagram showing the Discrete Wavelet Transform. Low pass and high pass filters are applied to the signal at each level, outputting respectively Approximation components $\textbf{Ak}$ and Detail components $\textbf{Dk}$.}
\label{fig:fig2}
\end{center}
\end{figure}

\begin{figure}
\raggedright
\includegraphics[width=.95\columnwidth]{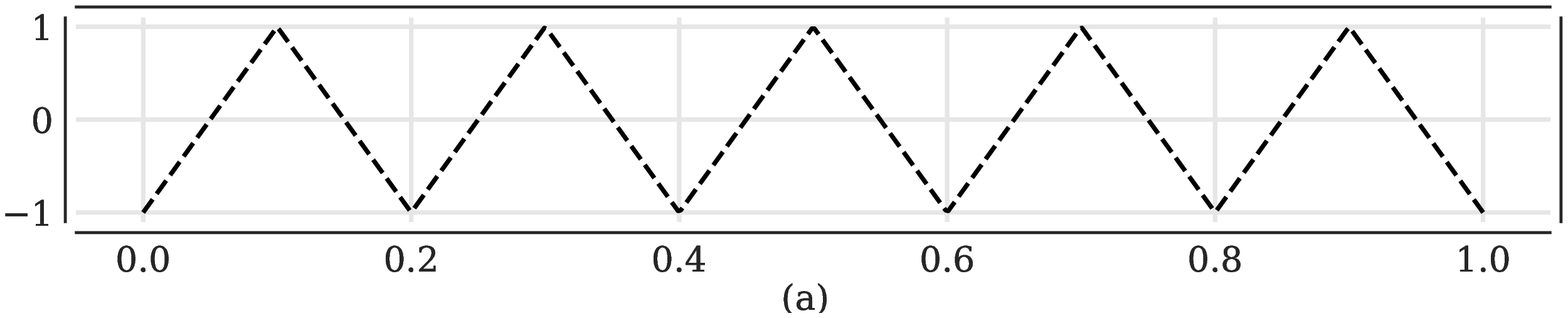}
\bigbreak
\includegraphics[width=.95\columnwidth]{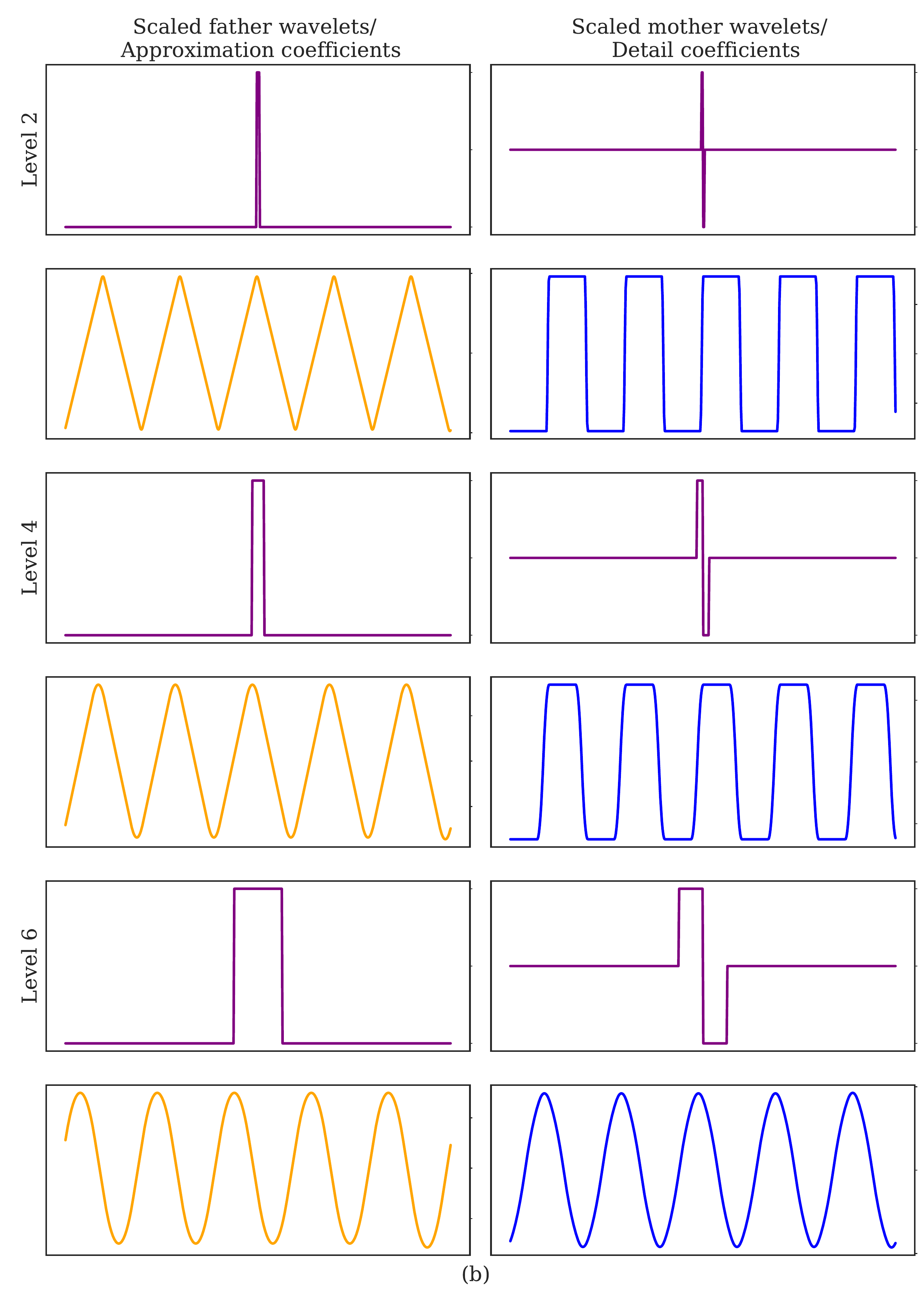}
\caption{The Discrete Wavelet Transform in action, using the `Haar' wavelet. {\it (a) Top Panel}---Input signal upon which DWT acts. The horizontal axis is time while the vertical axis is amplitude. There are 512 points in time; thus the maximum level of decomposition is 9 ($\log_2 512$). For illustration purposes, we have only shown the second, fourth, and sixth decomposition coefficients and wavelets. {\it (b)} Each six--panel column shows three DWT coefficients, along with the scaled father and mother wavelets at the relevant scale. The central tick on the vertical axis corresponds to 0. {\it Left Column}---Scaled father wavelets (purple) and Approximation coefficients (orange) of the input signal. {\it Right Column}---Scaled mother wavelets (purple) and Detail coefficients (blue) of the input signal.}
\label{fig:dwt_zigzag}
%\end{center}
%\end{subfigure}
\end{figure}

With increasing decomposition level, less information will be included in the approximation coefficients. The lost information between approximation coefficients of two successive decompositions is encoded in the detail coefficients. This process can be iterated to level $\log_2 N$, where the length of the original input is $N$. We must have $N$ a power of $2$ because the DWT is a recursive splitting. As a result, a vector of approximation coefficients and a series of vectors of detail coefficients are obtained that forms the DWT coefficients. The signal can be reconstructed by inverse DWT. Fig. \ref{fig:fig2} presents a block diagram demonstrating this concept. Fig. \ref{fig:dwt_zigzag} demonstrates the DWT process when using a `Haar' wavelet on a sawtooth signal. As is demonstrated there, at each successive level (scale), the mother and father wavelets are dilated by a factor of 2, and convolved with the Approximation coefficient from the previous level. Since the father wavelet is non--negative, this is equivalent to adding successive terms of the input signal (i.e., the Approximation coefficient from the previous level) in the time domain to get the Approximation coefficient for the next level. Similarly, the action of the mother wavelet is to find differences between pairs of points for all points in the time domain; this is how we obtain the Detail coefficient for the next level. These dual actions of scaling up (equivalent to increasing the window-size of the kernel in Fourier transform) and convolving with the mother and father wavelets help DWT achieve finite resolutions in both time and frequency domains---a significant advantage over the Fourier transform.

%\afterpage{\FloatBarrier}

\subsection{Shrinkage}\label{shrinkage}
DWT can be used for easy and fast denoising of a noisy signal. If we take only a limited number of highest approximation coefficients of the discrete wavelet transform spectrum and perform an inverse transform, we can obtain a partially denoised signal; this process is known as \emph{wavelet shrinkage}. A more nuanced approached involves \emph{thresholding} the values of the approximate coefficients that lie above a certain level (the \emph{threshold}), instead of rejecting coefficients entirely \citep{johnstone2005empirical}.

There are several different thresholding schemes in the literature, e.g. \citet{johnstone2005empirical},  \citet{li2007wavelet}. The most common two are soft and hard thresholding. In soft thresholding, the $i^{th}$ component of the $j^{th}$ coefficient, $w_{j,i}$, will be set to zero if it is smaller than the threshold for that level $\lambda_{j}$. We note that $j$ runs from $0$ up to the maximal decomposition level $k$, and we use just one threshold value at each level $j$. Otherwise, the $w_{j,i}$ are reduced by $\lambda_{j}$. With $\widetilde{w}_{j,i}$ denoting the thresholded coefficient, we have
\begin{equation} \label{eq:eq9}
\widetilde w_{j,i}^{SOFT} =
\begin{cases}
%\begin{align*}
&\text{sgn}(w_{j,i})(|w_{j,i}| - \lambda_{j}), \;\;\;{\rm for}\;|w_{j,i}| \geq \lambda_{j}\\    
&0, \;\;\;{\rm for}\;|w_{j,i}| < \lambda_{j}.  
%\end{align*}
\end{cases}
\end{equation} \\
In hard thresholding, the coefficients that are smaller than the threshold value are set to zero; those equal to it or larger are not altered. We have
\begin{equation} \label{eq:eq10}
\widetilde w_{j,i}^{HARD} =
\begin{cases}
%\begin{align*}
&w_{j,i}, \;\;\;{\rm for}\; |w_{j,i}| \geq \lambda_{j}\\    
&0, \;\;\;{\rm for}\; |w_{j,i}| < \lambda_{j}.  
%\end{align*}
\end{cases}
\end{equation} 
The noise threshold $\lambda_j$ for a given Detail component can be obtained either using the universal level--independent threshold \citep{donoho1994ideal} $\lambda = \sigma^{\rm Noise}\sqrt{2\log{N}}$, with $\sigma^{\rm Noise}$ an estimate of the noise and $N$ the length of the input signal, or a level-dependent threshold such as in \citet{johnstone1997wavelet}, which sets $\lambda_{j} = \sigma_j^{\rm Noise}\sqrt{2\log_{2}N_{j}}$
%or $\lambda_{j} = \frac{ \sigma_j^{\widetilde{Noise}}\sqrt{2\log(N_{j})}}{log(j+1)}$
, where $N_j$ is the length of the $j^{th}$ Detail component, and $\sigma_j^{\rm Noise}$ is an estimate of the noise level.

\begin{figure}
\centering
\includegraphics[width=1\columnwidth]{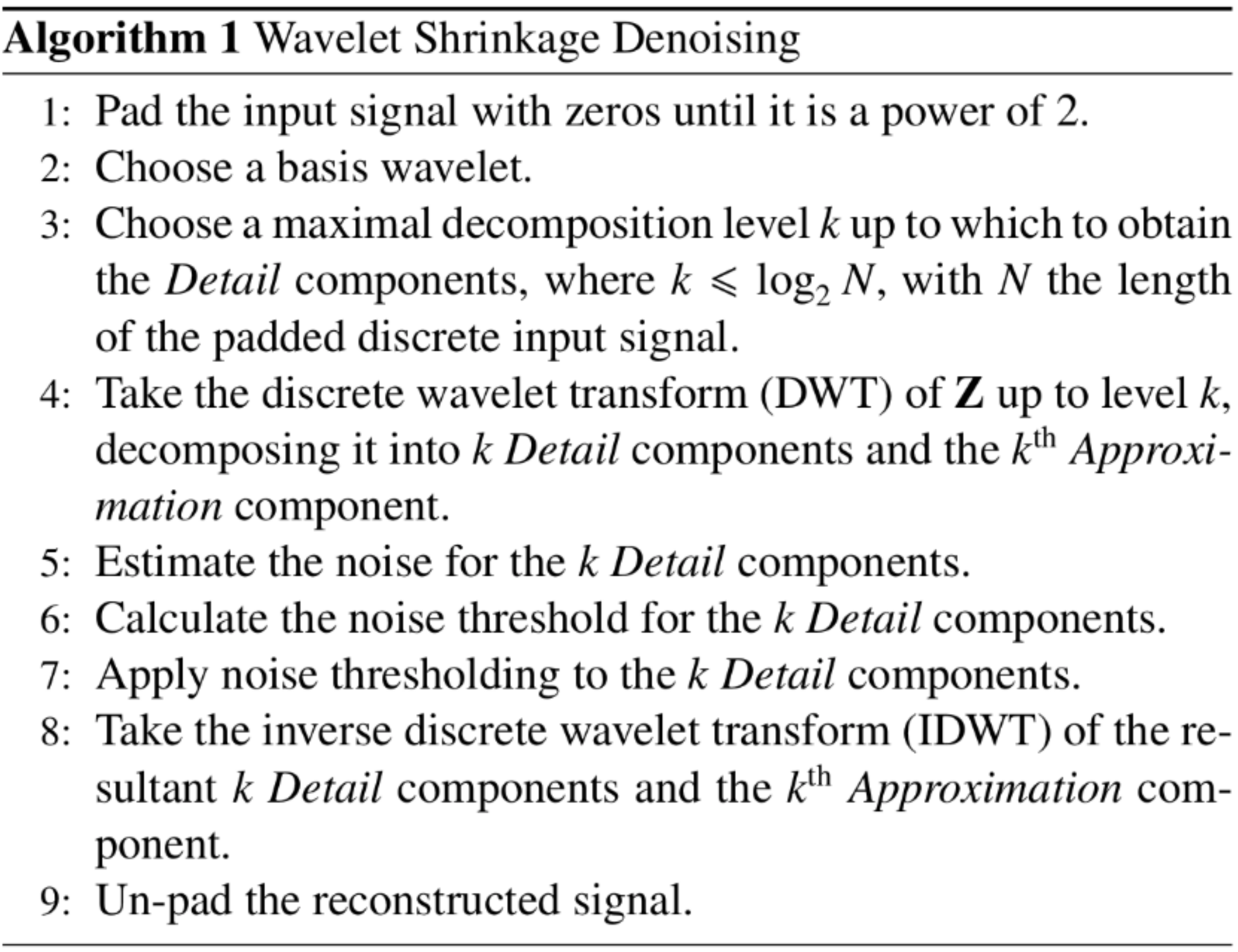}
\caption{Steps for the `traditional' wavelet shrinkage denoising method. The performance is highly dependent on the number of decomposition levels of the input signal, which the user is required to choose. Too shallow a decomposition, and not all the noise is removed; too deep a decomposition, and we might end up removing parts of the signal of interest.}
\label{algo1}
\end{figure}

\begin{figure}
\begin{center}
\includegraphics[width=1\columnwidth]{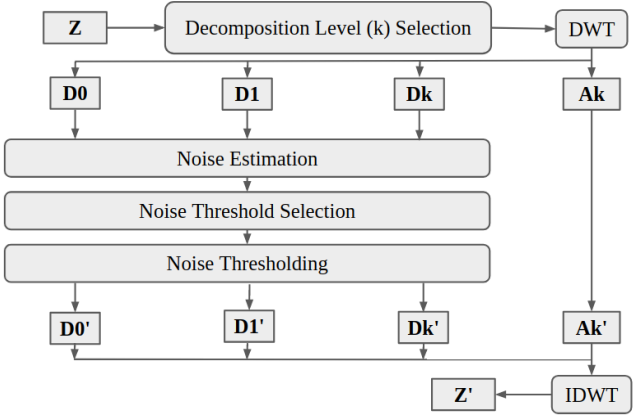}
\caption{Block diagram of Wavelet Shrinkage Denoising. The input signal $\textbf{Z}$ is decomposed into $k$ levels through the Discrete Wavelet Transform (DWT), to obtain $k$ Detail components $\textbf{D0}$-$\textbf{Dk}$, and one Approximation component $\textbf{Ak}$ (second line of figure). After selecting a noise level and a thresholding scheme (third and fourth line), the latter is applied to remove the noise from the Detail components. These thresholded Detail components {\bf D0\textquotesingle}-{\bf Dk\textquotesingle}, along with the unmodified Approximation component {\bf Ak} (= {\bf Ak\textquotesingle}) (all in second-to-last line of figure), are then converted back to a denoised signal {\bf Z\textquotesingle} via the Inverse Discrete Wavelet Transform (IDWT).}

\label{fig:fig3}
\end{center}
\end{figure}

Hard thresholding is better suited for the case where there is a Detail wavelet coefficient is either a signal or a noise coefficient. On the other hand, soft thresholding performs better when a Detail wavelet coefficient contains both signal and noise. While there are several thresholding functions in the literature that can be used \citep{chang2000adaptive, zhao2015improved, zhang2003denoising}, in this work, we have used the universal, hard, and soft thresholding schemes to compare to our proposed algorithm. Fig. \ref{algo1} summarizes the steps involved in the denoising process, while Fig. \ref{fig:fig3} displays a block diagram of the same.

A detailed explanation and description of wavelet-based methods can be found in \citet{graps1995introduction, taswell2000and, li2005wavelet, tary2018analysis}.

\section{Kalman Filter}\label{kalman}

%The Kalman filter is a set of mathematical equations that implement a predictor-corrector type estimator that is optimal in the sense that it minimizes the estimated error covariance—when some presumed conditions are met. Since the time of its introduction, the Kalman filter has been the subject of extensive research and application, particularly in the area of autonomous or assisted navigation. This is likely due in large part to advances in digital computing that made the use of the filter practical, but also to the relative simplicity and robust nature of the filter itself. Rarely do the conditions necessary for optimality actually exist, and yet the filter apparently works well for many applications in spite of this situation.

%The Kalman filter is named after Rudolph E. Kalman, who in 1960 published his famous paper describing a recursive solution to the discrete-data linear filtering problem (Kalman 1960). 
The Kalman filter is a method for updating predictions in the presence of noisy measurements. Named for its inventor Rudolph E. Kalman \citep{kalman1960new}, it uses a predictor-corrector system to recursively improve the accuracy of predictions by accounting for the noise properties of both the underlying process and the measurements. It has been extensively applied in many areas including control systems, tracking, navigation, and hydrology \citep{10.1002/9780470377819.ch1, yeh2005parameter}. Typically, the Kalman filter (and its advanced variants, such as the Unscented Kalman filter) are used to estimate the state of a time-varying process---such as tracking the location of an airplane using radar data and other relevant observables.

However, in this work we have used Kalman filtering to de-noise stellar absorption spectra by re-interpreting the time domain as the wavelength domain. Just as with time, points in wavelength follow a fixed ordering. However, unlike with time, wavelength does not have a fixed overall direction for correlations. Whereas by causality, a given timestep of a process should correlate only with previous timesteps, the value of a spectrum's flux on a bin in wavelength is likely correlated with both lower-wavelength and higher-wavelength bins, as spectra are generally roughly predictable. Our problem is thus a more symmetric one than the usual cases where Kalman filtering is applied. To preserve this symmetry, we apply Kalman filtering in both backward and forward wavelength directions while denoising the spectra, as detailed in \S\ref{main}. For an intuitive derivation of the linear Kalman filter, we refer the readers to Appendix \ref{kalman_derivation}.

\subsection{Modeling the Process}\label{kalman_implemention}
Let us denote the measured, noisy signal at wavelength bin \emph{i} by $\textbf{z}_{i}$, and the true, unknown signal by $\textbf{x}_{i}$. Let the process noise, i.e. the intrinsic noise describing the stochasticity of the system itself, independent of measurement, be $\textbf{w}_{i}$. Let the measurement noise---arising due to the in-practice imperfect process of measurement---be denoted by $\textbf{v}_{i}$. 

The Kalman filter implementation used in this paper assumes $\textbf{w}_{i}$ and $\textbf{v}_{i}$ are normally-distributed random variables with independent respective covariance matrices $\textbf{Q}_{i}$ and $\textbf{R}_{i}$. Since the Kalman filter is a recursive filter, the true (clean) signal at wavelength \emph{i} depends on the signal at wavelength $i-1$; let the evolution operator from $i-1$ to $i$ be denoted by $\textbf{M}_{i}$. Finally, let the operator relating the measured signal $\textbf{z}_{i}$ to the true signal $\textbf{x}_{i}$ be denoted by $\textbf{H}_{i}$. Then for any wavelength point \emph{i} the true signal (`state', hereafter) and the measurement signal satisfy the process equation (\ref{eq:eq1}) and the measurement equation (\ref{eq:eq2}), respectively:
\begin{align} \label{eq:eq1}
%\textbf{x}_{k} &= \textbf{A}_{k}\textbf{x}_{k-1} + \textbf{B}_{k}\textbf{u}_{k} + \textbf{w}_{k} \\
\textbf{x}_{i} &= \textbf{M}_{i}\textbf{x}_{i-1} + \textbf{w}_{i} \\
\label{eq:eq2}
\textbf{z}_{i} &= \textbf{H}_{i}\textbf{x}_{i} + \textbf{v}_{i}.
\end{align} 

%The Kalman filter is a mathematical tool for stochastic estimation of the state of a linear system from noisy measurements of the system. In other words, it is used to estimate the state of a linear dynamical system based on information from a measurement \citep{10.1002/9780470377819.ch1}. %It was proposed by R.E. Kalman in a seminal paper \citep{kalman1960new} to solve the discrete-data linear filtering problem from the system state point of view. It has been applied in many areas including control systems, tracking, navigation, and hydrology \citep{yeh2005parameter}. Below, we explain the functioning of the Kalman Filter in terms of temporal evolution, since this is what it was designed to solve, but we remind the reader that

Now that we have introduced the main elements of the Kalman filter, we proceed to briefly describe the set of recursive operations that estimate the clean signal at each wavelength \emph{i}. This is achieved in two steps \citep{welch2006introduction}: `prediction' and `update', as detailed in \S\ref{kalman_prediction} and \S\ref{kalman_update}. We outline the Kalman filtering process assuming we are moving in the direction of increasing wavelength; however as discussed earlier (\S\ref{kalman}), we also apply our process backwards. The appropriate equations can simply be obtained by switching $i-1$ and $i$. 

%\ref{eq:eq3} - \ref{eq:eq4}, and \ref{eq:eq5} - \ref{eq:eq7}, respectively \citep{welch2006introduction}.\\ \\
%Predict phase:
%\begin{align} \label{eq:eq3}
%\textbf{\^{x}}_{k}^{-} &= \textbf{A}_{k}\textbf{\^{x}}_{k-1} + %\textbf{B}_{k}\textbf{u}_{k} \\
%\label{eq:eq4}
%\textbf{P}_{k}^{-} &= %\textbf{A}_{k}\textbf{P}_{k-1}\textbf{A}_{k}^{T} + \textbf{Q}_{k}
%\end{align}
%Update phase:
%\begin{align} \label{eq:eq5}
%\textbf{K}_{k} &= \textbf{P}_{k}^{-}\textbf{H}_{k}^{T}(\textbf{H}_{k}\textbf{P}_{k}^{-}\textbf{H}_{k}^{T} + \textbf{R}_{k})^{-1} \\
%\label{eq:eq6}
%\textbf{\^{x}}_{k} &= \textbf{\^{x}}_{k}^{-} + \textbf{K}_{k}(\textbf{z}_{k} - \textbf{H}_{k}\textbf{\^{x}}_{k}^{-}) \\
%\label{eq:eq7}
%\textbf{P}_{k} &= (\textbf{I} - \textbf{K}_{k}\textbf{H}_{k})\textbf{P}_{k}^{-}
%\end{align}
%Here $\textbf{\^{x}}_{k}^{-} \epsilon R^{n}$ is the \textit{a priori} state estimation at time \textit{k}, the $n$x$n$ matrices $\textbf{P}_{k}^{-}$ and $\textbf{P}_{k}$ are the \textit{a priori} and the \textit{a posteriori} estimation error covariance, respectively, and the $n \times l$ matrix $\textbf{K}_{k}$ is the Kalman gain.
\subsection{Predicting the Process}\label{kalman_prediction}
In the `predict' phase, the Kalman filter uses information from the previous state to estimate the \textit{a priori} current state. This estimate is then updated using information from a measurement to produce an \textit{a posteriori} state estimate: this is termed the `update' phase.

In more detail, first we initialize a mean ($\textbf{\^{x}}_{-1}$) and a covariance matrix ($\textbf{P}_{-1}$) for the state of the system, which are advanced to obtain an \emph{a priori} estimate of the state (mean $\textbf{\^{x}}_{0}$ and covariance $\textbf{P}_{0}$). The covariance part is necessary to handle the uncertainty in the estimate arising due to the failure of the linear model used for the system. This is the `prediction' part of the Kalman filtering process---where we use the estimate of the system state from the previous run of the process to predict the \emph{a priori} for the next run; i.e., use the flux predicted for wavelength \emph{i}-1 to get an \emph{a priori} estimate for the flux at wavelength \emph{i}. This is encoded in equations (\ref{eq:eq3}) and (\ref{eq:eq4}):
%\textsc{Predict phase}:
\begin{align} \label{eq:eq3}
\textbf{\^{x}}_{i}^{-} &= \textbf{M}_{i}\textbf{\^{x}}_{i-1} \\ %+ \textbf{B}_{i}\textbf{u}_{i} \\
\label{eq:eq4}
\textbf{P}_{i}^{-} &= \textbf{M}_{i}\textbf{P}_{i-1}\textbf{M}_{i}^{T} + \textbf{Q}_{i}.
\end{align}

\subsection{Updating the Process}\label{kalman_update}
Next, we use the measured signal at wavelength \emph{i} ($\textbf{z}_{0}$, since $i = 0$ in the first iteration) to update the \emph{a priori} state prediction from the previous step (i.e., $\textbf{\^{x}}_{0}^{-}$ and $\textbf{P}_{0}^{-}$). This measurement comes from an instrument (a spectrograph in our case); its uncertainty is the measurement noise ($\textbf{v}_{i}$ at wavelength \emph{i}, with assumed mean $0$ and covariance matrix $\textbf{R}_{i}$). This is induced by inaccuracies in the measurement process (e.g. an imperfect CCD chip or an unstable spectrograph). In addition, we class within this measurement error any errors arising due to imperfect modeling of the system from our assumption that its evolution is linear; we highlight in \S\ref{main} our rationale for this choice.%Given these two state estimates and their related uncertainties, under specific assumptions, the Kalman filter combines the \emph{a priori} state estimate and the measurement to generate an \emph{a posteriori} state estimate, whose uncertainty is minimized.
 This phase is known as the `correction' or the `update' phase; it is encoded in equations (\ref{eq:eq5}), (\ref{eq:eq6}) and (\ref{eq:eq7}):
%\textsc{Update phase:}
\begin{align} \label{eq:eq5}
\textbf{K}_{i} &= \textbf{P}_{i}^{-}\textbf{H}_{i}^{T}(\textbf{H}_{i}\textbf{P}_{i}^{-}\textbf{H}_{i}^{T} + \textbf{R}_{i})^{-1}, \\
\label{eq:eq6}
\textbf{\^{x}}_{i} &= \textbf{\^{x}}_{i}^{-} + \textbf{K}_{i}(\textbf{z}_{i} - \textbf{H}_{i}\textbf{\^{x}}_{i}^{-}), \\
\label{eq:eq7}
\textbf{P}_{i} &= (\textbf{I} - \textbf{K}_{i}\textbf{H}_{i})\textbf{P}_{i}^{-}.
\end{align}
In equation (\ref{eq:eq5}), $\textbf{K}_{i}$ refers to the Kalman gain, which encodes the filter's confidence in the prediction $\textbf{\^{x}}_{i}^{-}$ versus in the measurement $\textbf{z}_{i}$. As one might expect, the Kalman gain depends on the error covariance matrices associated with both these quantities. In the limiting case where the measurement error at a wavelength point \emph{i} is much smaller than the prediction error (i.e., $\textbf{R}_{i} \ll \textbf{P}_{i}^{-}$), $\textbf{K}_{i} \to \textbf{H}_{i}^{-1}$ and the \emph{a posteriori} state matches the measurement as acted upon by the inverse of the measurement matrix (i.e, $\textbf{\^{x}}_{i} \to \textbf{H}_{i}^{-1}\textbf{z}_{i}$ and  $\textbf{P}_{i} \to 0$). Since the present work deals with denoising spectra as measured, which are obtained by spectrographs, both the true state $\textbf{\^{x}}$ and the measurement $\textbf{z}$ refer to spectra. Hence all $\textbf{H}s$ have been assigned the identity matrix, and the \emph{a posteriori} state matches the measurement exactly ($\textbf{\^{x}}_{i} \to \textbf{z}_{i}$). This makes intuitive sense---when we are extremely confident about a measurement, we can assume that to be the real state of the system. On the other hand, when the measurement error dominates (i.e., $\textbf{R}_{i} \gg \textbf{P}_{i}^{-}$), $\textbf{K}_{i} \to 0$ and the \emph{a posteriori} state matches the \emph{a priori} state (i.e, $\textbf{\^{x}}_{i} \to \textbf{\^{x}}_{i}^{-}$ and  $\textbf{P}_{i} \to \textbf{P}_{i}^{-}$). When we lack confidence in a measurement, it is prudent not to update the original \emph{a priori} state estimate. In equation (\ref{eq:eq7}), $\textbf{I}$ refers to the identity matrix.

This entire process is repeated recursively until we have traversed the length of the input signal. As noted earlier, to encode the symmetry under traversing the wavelength range forwards or backwards, we run our spectral denoising algorithm for each traversal direction.

\section{Limitations of Current Methods}\label{limitations}
The standard soft-- and hard--thresholding functions, along with other commonly used threshold-selection schemes, suffer from a few major drawbacks that preclude their use for automated signal denoising:
\begin{enumerate}
\item The choice of $k$, the number of wavelet transform decomposition levels to threshold, is data-dependent and needs to be carefully chosen by the user.
\item The choice of $\sigma_j^{\rm Noise}$, the noise estimate at the $j^{\rm th}$ decomposition level, greatly influences the noise threshold $\lambda_j$, but there is no definitive way to estimate it. Different noise estimates yield different noise thresholds \citep{donoho1994ideal, donoho1995adapting, johnstone1997wavelet, srivastava2016new}.
\item Assuming white Gaussian noise (WGN), a single noise threshold is selected and applied to the magnitudes of both negative and positive Detail coefficients. However for WGN, the assumption of no distribution bias (i.e. a distribution symmetric about zero) may often be flawed \citep{srivastava2016new}.
\item The methods used result in fixed thresholds that are not adjusted. Flexibility in adjusting thresholds is extremely important to optimally threshold for signals, especially when the wavelet coefficients of weak signals are close to the maximum magnitude of noise \citep{srivastava2016new}.
\end{enumerate}

The method proposed in the current work is considerably more `non-parametric' than both the traditional wavelet denoising schemes discussed above, and traditional smoothing or filtering schemes like the Fourier transform, moving mean, etc. The only `parameter' of consequence that our proposed algorithm requires from the user is a suitable wavelet function. The criteria for choosing this basis function are: maximization of its correlation with the signal, and minimization of its correlation with the noise. Clearly, we do not want to choose an arbitrary wavelet function. The selection and development of such data-specific basis wavelets is well-studied (e.g. \citealt{Rafiee:2009:NTS:1497653.1498474}). In the current work, we use the simplest and computationally cheapest wavelet (`Haar').
%During wavelet-based de-noising, we need to consider two issues: wavelet basis function selection and setting of thresholds.In the application of wavelet-based de-noising, the selection of a suitably wavelet function is an important problem. The criteria of choosing a basis function are the maximization of its correlation with the information signal, as well as the minimization of its correlation with the noisy signal; clearly, we do not want to choose an arbitrary wavelet function. The selection and development of such data-specific `mother' wavelets is a well researched field, and we refer the interested reader to peruse them [].
%Several factors orthogonality, wavelets shape and wavelet width should be considered. Orthogonal shape and wavelet width should be considered. Orthogonal wavelet can give the most conpact repressenttion of a signal, so this study concentrates on orthogonal wavelet functions. Daubechies wavelet has good localizing properties both in temporal and frequency domains, however bi-orthogonal wavelet has symmetry and simplicityIf a signal consists of many sharp jumps or stepsm a boxcar-like function such as the Haar wavelet is reasonable, while for smoothly varying time series a smooth function such as Symlets wavelet and discrete approximation of Meyer wavelet asre better.

\section{Automatic Stellar Denoising}\label{main}
In \S\ref{test_data}, we describe the nature and origin of the stellar spectra used to compare the performance of the proposed method with the major denoising methods from the literature. In \S\ref{akft}, we explain in detail the implementation of two versions of our proposed algorithm. We then introduce error metrics used to compare the various filtering algorithms, and end by demonstrating the importance of the non-parametric nature of our proposed algorithm.

\subsection{Test data}\label{test_data}
We use synthetic 1D PHOENIX spectra \citep{husser2013new} for three stars of different spectral types. \emph{Star 1} has T$_{\rm eff} = 4,000$K, [Fe/H] = $1.0$ dex, $\log g = 2.0$ dex, and spectral type = KIII. \emph{Star 2} has T$_{\rm eff} = 6,000$K, [Fe/H] = $1.0$ dex, $\log g = 3.5$ dex, and spectral type = GV. \emph{Star 3} has T$_{\rm eff} = 7,000$K, [Fe/H] = $1.0$ dex, $\log g = 2.5$ dex, and spectral type = KVI. The spectra are convolved from the native PHOENIX resolution of $R = \lambda/\Delta\lambda = 500,000$ down to the three lower resolutions of $2,000$, $5,000$ and $10,000$, which are then re-sampled onto the optical wavelength interval of $5,000 - 5,495.9$\si{\angstrom}, with a sampling of $0.1$\si{\angstrom}. The wavelength range and the sampling chosen are for illustration purposes and have no bearing on the generalizability of our proposed algorithm. Our sampling yields vectors of length $4,096 = 2^{12}$. Each of the nine spectra were disturbed by twenty-five different levels of Poisson noise, to obtain final PSNRs (peak signal--to--noise ratios) %, defined as the square root of the maximum signal amplitude)
going in steps of unity from $1.0-25.0$. For each spectrum and PSNR, we produced $100$ realizations; this point is further discussed in \S\ref{results}. Both the  \emph{reference} spectra and the corresponding \emph{noisy} spectra were normalized by the peak flux in the \emph{reference} spectra to make them all lie on a comparable flux scale. In total we deal with $9\times 25 \times 100 = 22,500$ unique vectors to denoise, each consisting of $4,096$ data points. 

\subsection{Adaptive Kalman-Filter-based Thresholding (AKFThresh)}\label{akft}
As described in Section \ref{limitations}, denoising performance with conventional wavelet shrinkage methods is a strong function of the number of decomposition levels and the thresholding scheme. Our proposed algorithm deals with this dependence by treating the individual decomposed Detail components of the noisy input signal as separate signals to denoise, and utilizes distinct Kalman filters (see equations \ref{eq:eq3}-\ref{eq:eq7}) for each. 
%Sankalp: update equation reference if need be!

In brief, our proposed thresholding scheme works as follows. With every Detail level $j$, from $0\leq j \leq k$, we associate a Kalman filter. At this level, the noisy signal $\textbf{Z}_{Dj}$ is the Detail component, the denoised signal is $\textbf{X}_{Dj}$, and $\textbf{H}_{Dj}$ is set to $\textbf{I}$ (the identity matrix). We are thus tasked with obtaining $\textbf{Q}_{Dj}$, $\textbf{R}_{Dj}$, and $\textbf{M}_{Dj}$ (see equations \ref{eq:eq3}-\ref{eq:eq7}). 
%Sankalp: update equation references if need be.

Since the $0^{\rm th}$ level Detail component consists of the highest-frequency part of the spectral signal, we assume it to be completely noise; specifically, we set the process noise covariance vector for the $0^{\rm th}$ Detail component $\textbf{Q}_{D0}$ to $\textbf{0}$, and use the algorithm suggested by \citet{meng2016covariance} to obtain the measurement noise covariance matrix  $\textbf{R}_{D0}$. Fig. \ref{fig:dwt_snr10_R0} illustrates the importance of choosing $\textbf{R}_{D0}$ in this manner. Starting from the $1^{\rm st}$ level Detail component, all the measurement noise covariance matrices $\textbf{R}_{Dj}$ are set equal to $\textbf{R}_{D0}$, while the process noise covariance matrices $\textbf{Q}_{Dj}$ are obtained by subtracting the variance of the $0^{\rm th}$ Detail component from the covariance of the $0^{\rm th}$ and the $j^{th}$ Detail components. Even though we are dealing with synthetically-generated spectra in the absence of a measurement device, the measurement noise matrices $\textbf{R}_{Dj}$ are not {\bf 0} since they capture the errors emanating from the imperfect assumption of linearity in our Kalman filters. 

%Sankalp: please explain the R matrix thing!
At the same time, we estimate the state transition matrix ${\textbf{M}_{Dj}}$ by following the Support Vector Regression (SVR) procedure outlined in \citet{salti2013line} rather than setting it to unity. It is precisely this heuristic of obtaining $\textbf{R}_{Dj}$, $\textbf{Q}_{Dj}$, and $\textbf{M}_{Dj}$ for the Kalman filters at the various Detail levels that allows us to make the process effectively non-parametric, instead of finding the optimal number of decompositions as is done traditionally. The pseudo-code for the proposed method is given in Fig. \ref{algo2}. All the variances and covariances are obtained in a robust fashion.\footnote{\url{https://scikit-learn.org/stable/modules/generated/sklearn.covariance.MinCovDet.html}}

We use two versions of our denoising algorithm---\textsc{AKFThresh v1} refers to the algorithm outlined above (and in Fig. \ref{algo2}), whereas \textsc{AKFThresh v2} uses the denoised output vector from \textsc{AKFThresh v1} as input, takes the natural logarithm of the values (after adding an offset to make them strictly positive, if required), applies the \textsc{AKFThresh v1} algorithm again, and exponentiates the resulting vector (followed by removing any offset that was added). The Kalman filter assumes the process and measurement noises to be additive (see equations \ref{eq:eq3} and \ref{eq:eq4}). 
%Sankalp: update equations if need be.
Since Poisson noise---unlike Gaussian noise---is not strictly additive, this additional process is done to convert any residual multiplicative noise to additive form, and then remove it \citep{zhang2014method}.

We compare the denoising performance of the above two versions of the proposed algorithm with those of popular denoising algorithms in the field. These are: the moving mean,\footnotemark \;the moving median,\footnotemark[\value{footnote}] \footnotetext{\url{https://pandas.pydata.org/pandas-docs/stable/generated/pandas.Series.rolling.html}} exponential smoothing,\footnote{\url{https://pandas.pydata.org/pandas-docs/stable/generated/pandas.DataFrame.ewm.html}} Fast Fourier Transform with a Gaussian kernel,\footnote{\url{https://docs.scipy.org/doc/scipy/reference/generated/scipy.ndimage.gaussian_filter.html}} Wiener filter,\footnote{\url{https://docs.scipy.org/doc/scipy/reference/generated/scipy.signal.wiener.html}} the Savitzky--Golay (SG) filter,\footnote{\url{https://docs.scipy.org/doc/scipy-0.15.1/reference/generated/scipy.signal.savgol_filter.html}} wavelet transform followed by soft thresholding, and wavelet transform followed by hard thresholding. %(detailed discussion on WT, FT, S-G smoothing, and MA can be found in [18, 28, 29, 31].).
Since these commonly used methods are all parametric, %(besides the two versions of the algorithm proposed in this work, once the mother wavelet is fixed as `Haar')
any such comparison must be preceded by the selection of an appropriate loss function to minimize, followed by a search in their respective parameter spaces to optimize this loss function. Following \citet{kohler2005comparison} we calculate three loss functions---mean absolute error, or the $L^1$ norm, least squares error, or the $L^2$ norm, and maximum absolute error, or the $L^{\infty}$ norm. We also consider a fourth function, the structural similarity index measure or \emph{SSIM}, which is a measure of the similarity between two images \citep{wang2004image, wang2009mean}. 

\begin{figure}
\begin{center}
\includegraphics[width=1\columnwidth]{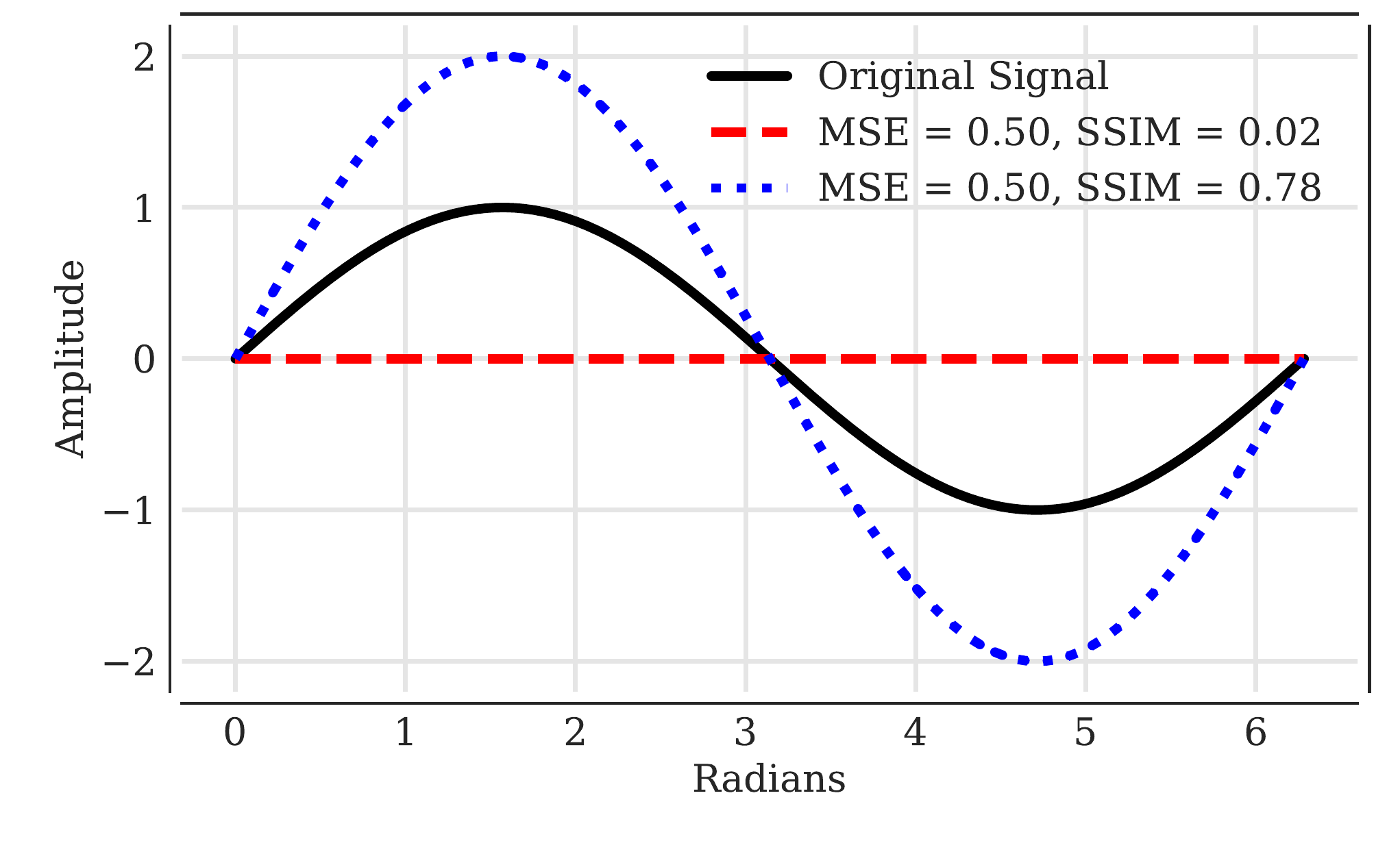}
\caption{Here we illustrate the difference between Mean Squared Error (MSE) and Structural Similarity Index Measure (SSIM) to emphasize the importance of assessing algorithm performance with respect to several different error metrics. We employed MSE in Fig. \ref{fig:fig6} and SSIM in Fig. \ref{fig:fig7}. The solid black curve represents the clean signal, while the dashed red and dotted blue curves represent two noisy versions of it. The MSE values for both noisy signals with respect to the clean signal are the same (0.50), while the SSIM values clearly illustrate that the blue signal (SSIM $= 0.78$) is much more similar to the original signal than is the red signal (SSIM $= 0.02$). This is of course also evident by visual inspection.}
\label{fig:mse_vs_ssim}
\end{center}
\end{figure}

\begin{figure*}
%\captionsetup[subfigure]{labelformat=empty}
%https://tex.stackexchange.com/questions/165508/remove-a-b-from-subfigure-numbering-but-keep-the-subfigure-caption
%second answer
%\begin{center}
\centering
%\begin{subfigure}{1\textwidth}
\includegraphics[width=.95\textwidth]{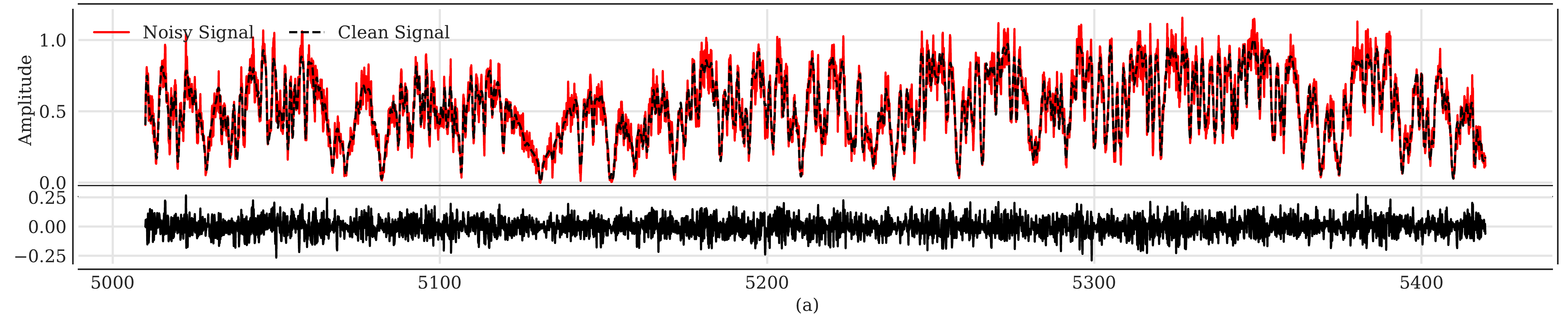}
\includegraphics[width=.95\textwidth]{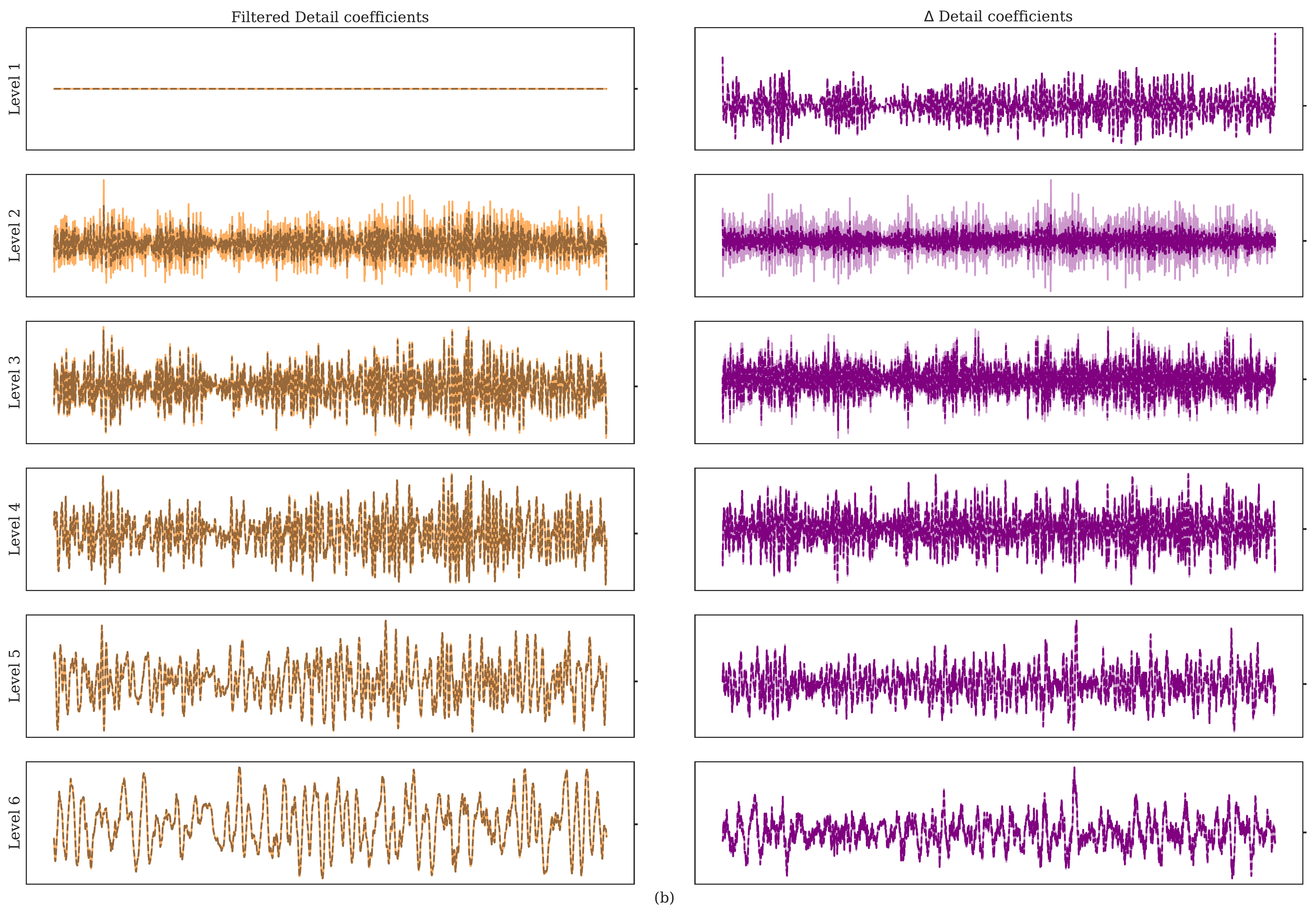}
\includegraphics[width=.95\textwidth]{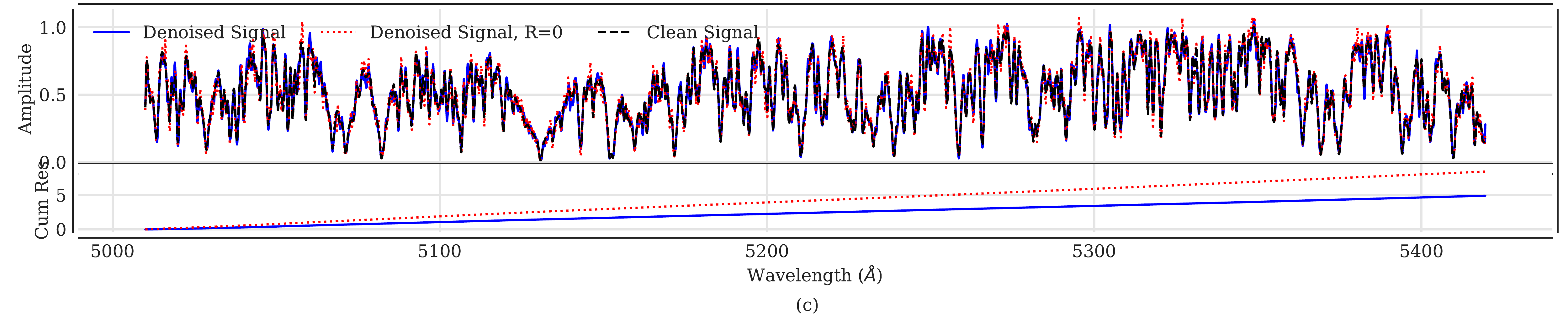}
\caption{Figure illustrating the importance of using a non-zero measurement covariance matrix $\textbf{R}$. {\it (a) Top Panel, upper sub-plot}---Clean (normalized) and noisy (peak signal-to-noise ratio $= 10$) synthetic stellar spectra for \emph{Star 1}. {\it Top Panel, lower sub-plot}---Residual between the clean and noisy signals. {\it (b)}---Each six--panel column shows the first six Discrete Wavelet Transform coefficients and adaptive Kalman filtering of the Detail coefficients. A horizontal line through the tick--mark of each corresponds to a vertical-axis value of $0$. {\it Left Column}---Denoised Detail coefficients obtained by applying Kalman filters to the noisy counterparts. Light orange represents values with the heuristically chosen non-zero $\textbf{R}$, while dark orange represents decomposition with this matrix set to $\textbf{0}$ throughout the wavelength range. {\it Right Column}---Residuals obtained by subtracting the Detail coefficients obtained from denoising using non-zero $\textbf{R}$, from the clean signal's Detail coefficients (solid lavender) and subtracting the same from the clean signal's Detail coefficients (double-dashed purple). {\it (c) Upper panel}---Normalized clean (dashed black), denoised with non-zero $\textbf{R}$ (solid blue), and denoised with $\textbf{R} = \textbf{0}$ (red) spectra for \emph{Star 1}. {\it Lower panel}--- Cumulative residual between the clean and the two versions of the denoised signals. It can be clearly seen that our scheme of choosing $\textbf{R}$ by following \citet{meng2016covariance} instead of forcing it to $\textbf{0}$ results in improved denoising performance.}
\label{fig:dwt_snr10_R0}
%\end{center}
%\end{subfigure}
\end{figure*}

\begin{figure}
%\centering
%\raggedright
%\includegraphics[width=1.8\columnwidth]{algo2_fixed}
\includegraphics[width=.9\textwidth,center]{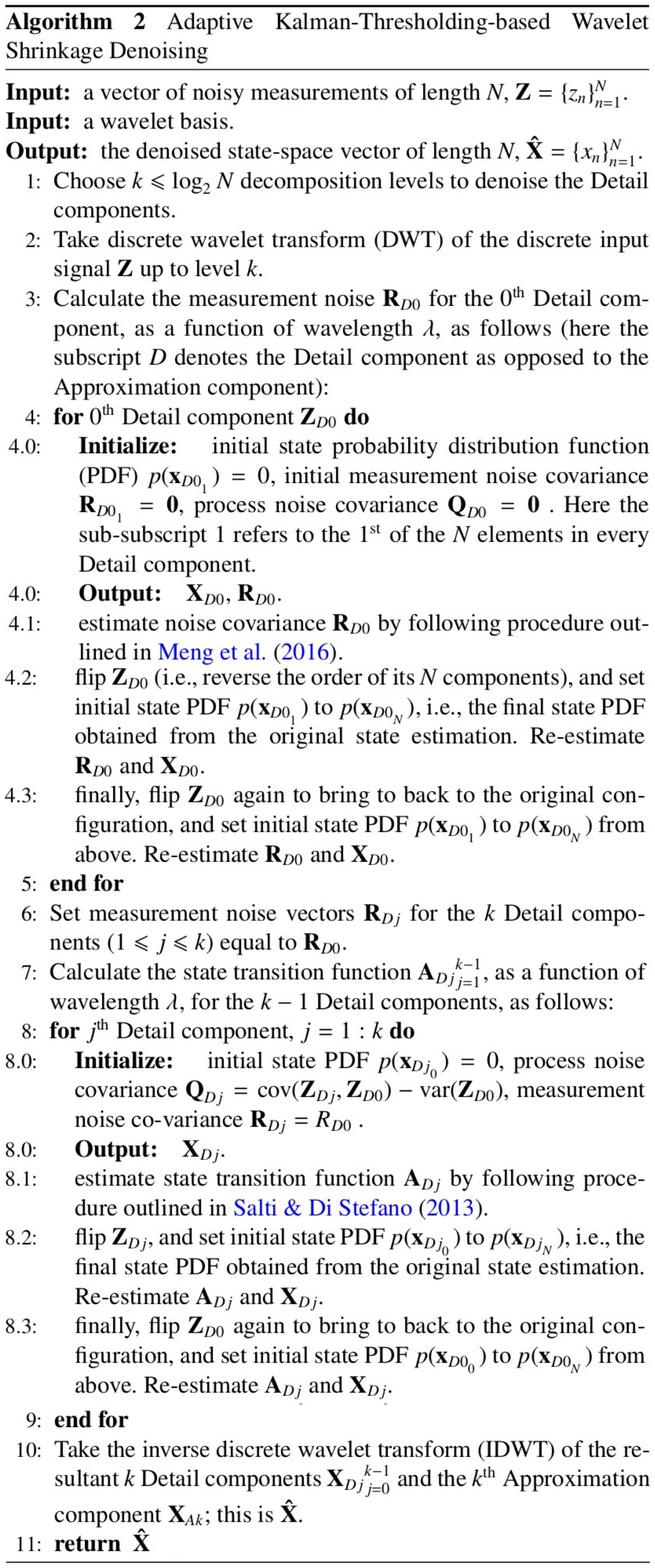}
\caption{Steps for the proposed adaptive Kalman-thresholding-based wavelet shrinkage denoising method. By adaptively setting the noise levels for the \emph{Detail} components instead of using a fixed level as done in Fig. \ref{algo1}, the denoising performance of the process becomes only weakly dependent on the number of decompositions. This is illustrated in Figs. \ref{fig:fig8} and \ref{fig:fig9}.}
\label{algo2}
\end{figure}

Each error metric probes the differences between given signals in a different way. We illustrate this with an example in Fig. \ref{fig:mse_vs_ssim}, by highlighting the differences between the mean squared error (MSE)---a `global' error metric---and the structural similarity index measure (SSIM)---a `local' error metric. We assume a sinusoidal form for the `clean' signal, and create two corrupted versions of it. The MSEs for the two versions of the `noisy' signal (dotted blue and dashed red curves) with respect to the real signal (solid black curve) are the same, while the SSIM is significantly higher for the dotted blue curve. This is also apparent by visual inspection---the dotted blue curve, while having a larger amplitude than the original signal, is clearly more `similar' to it, in terms of capturing the peak and the valley. In contrast, the dashed red curve, being identically zero, is devoid of any information about the original signal.

We should also optimize the parameters within each method with respect to the measure of quality we want to use. For instance, in a use case, if one wished to achieve minimum least-squares error on a given data set, clearly one would optimize one's parameters (if possible) under that error metric. We analyzed each norm to fully assess the strengths and weaknesses of the different denoising methods under each.

\begin{figure*}
%\captionsetup[subfigure]{labelformat=empty}
%https://tex.stackexchange.com/questions/165508/remove-a-b-from-subfigure-numbering-but-keep-the-subfigure-caption
%second answer
%\begin{center}
\centering
%\begin{subfigure}{1\textwidth}
\includegraphics[width=.95\textwidth]{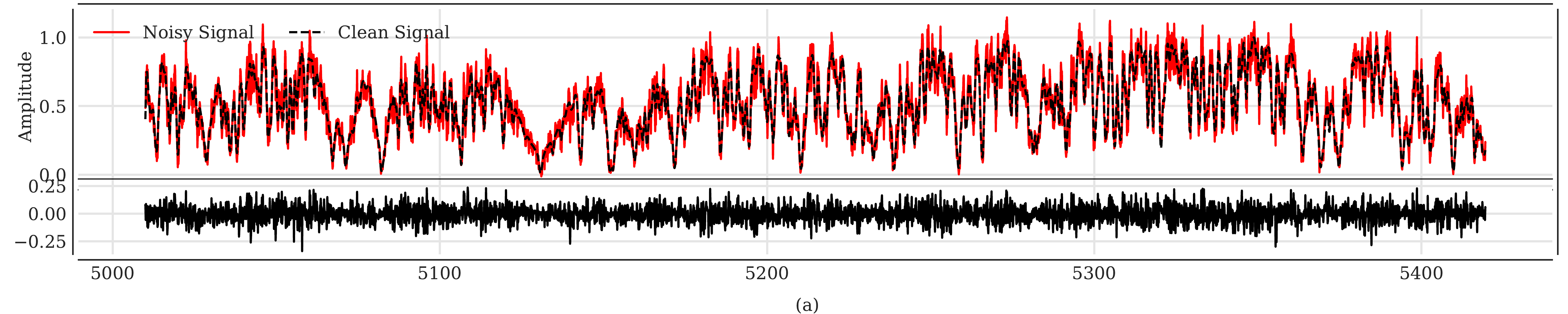}
\includegraphics[width=.95\textwidth]{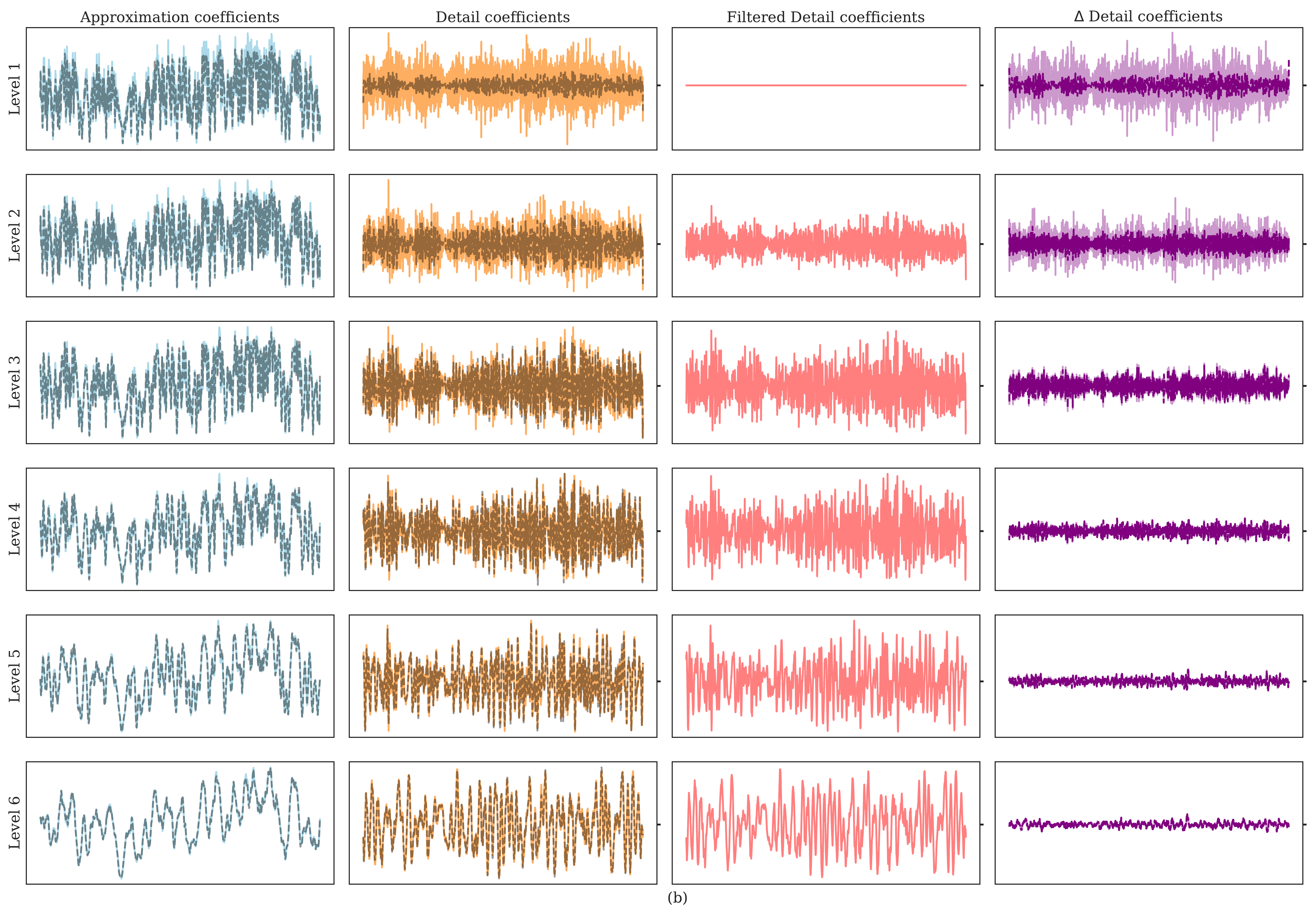}
\includegraphics[width=.95\textwidth]{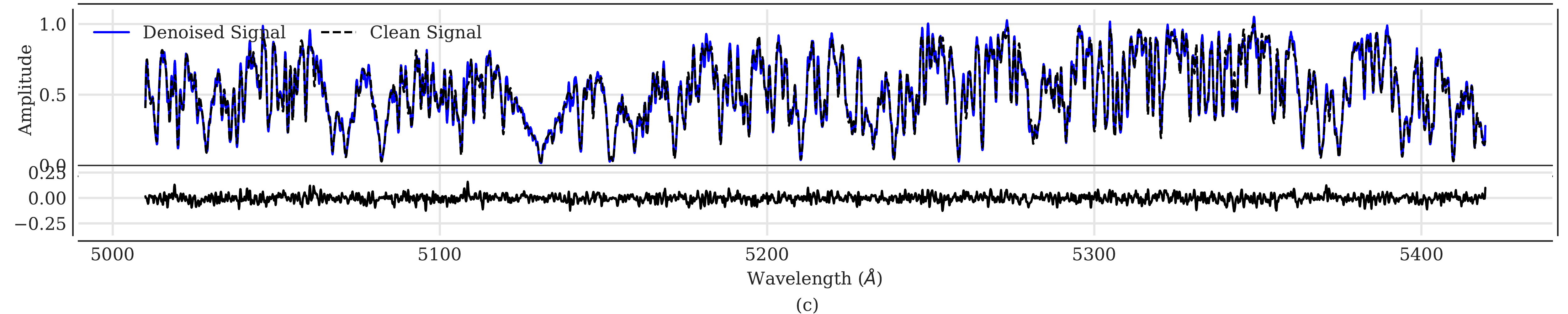}
\caption{This work's proposed signal denoising algorithm in action. {\it (a) Top Panel}---Exactly like in {\it (a)} of Fig. \ref{fig:dwt_snr10_R0}. {\it (b)}---The four columns of six panels on the right show the first six Discrete Wavelet Transform coefficients and adaptive Kalman filtering of the Detail coefficients. A horizontal line through the tick--mark of each corresponds to a vertical-axis value of $0$. {\it Leftmost Column}---Approximation coefficients for the clean (double-dashed, black) and noisy (solid, cyan) signals. {\it Second-from-left Column}---Detail coefficients for the clean and noisy signals; the clean Detail coefficient is in dashed brown, noisy in solid orange. The differences between the Detail coefficients for the clean and noisy signals are much larger than the differences between Approximations for clean (black, dashed) and noisy (light blue), thus justifying our decision to denoise Detail coefficients only. {\it Second-from-right Column}---Detail coefficients denoised by applying a Kalman filter to the Detail coefficients of the noisy signal (i.e. the orange curves in the left-neighboring column). {\it Rightmost Column}---Residuals obtained by subtracting the clean signal's Detail coefficients from the noisy signal's (solid lavender) and subtracting the clean signal's Detail coefficients from the denoised signal's (double-dashed purple). {\it (c) Top panel, upper sub-plot}---Normalized clean (dashed black) and denoised (solid blue) spectra for \emph{Star 1}. The denoised spectrum was obtained by applying an Inverse Discrete Wavelet Transform to the $6^{th}$ Approximation coefficient of the noisy signal (i.e the last row, leftmost panel in {\it (b)}) and the 6 denoised Detail coefficients (full second--from--right column in {\it (b)}). {\it Top panel, lower sub-plot}--- Residual between the clean and denoised signals. This is plotted on the same scale as the residual in the lower panel of {\it (a)}, to highlight the closer match of the denoised signal to the original signal.}
\label{fig:dwt_snr10}
%\end{center}
%\end{subfigure}
\end{figure*}

We define appropriately expansive parameter spaces for the different algorithms and do a thorough search to find the optimal parameters to minimize the $L^2$ norm between the denoised signal and the original reference signal. We repeat this analysis two more times, with the difference that now we look for parameters that give the tenth-lowest and the twentieth-lowest $L^2$ norms. This is done to simulate the situation where, in the absence of a reference signal, manual parameter search by trial-and-error yields sub--optimal parameters for a given denoising algorithm. We then compare the denoised signals from both versions of our proposed algorithm (\textsc{AKFThresh v1} and \textsc{AKFThresh v2}) to the denoised signals from the  \emph{optimal} and \emph{sub-optimal} versions of the other algorithms; partial tables displaying these results are available in Appendix \ref{nine_tables} (complete tables are available online only\footnote{\url{https://github.com/astrogilda/stellar_denoising}}). %Tables \ref{table:tab1} through \ref{table:tab3} display these results in full detail.
For robustness, this entire procedure was repeated for $100$ different realizations of the noise to create synthetic noisy signals for each of the three stars, and the results averaged. A high-level pictographic overview of our proposed method is presented in Fig. \ref{fig:dwt_snr10}, where we show its denoising performance in action for the absorption spectrum for \emph{Star 1} corrupted by Poisson noise to bring the PSNR down to $10$.
%Sankalp: did you mean 100 realization of the noise only, but with the same true signal? That's what I'm guessing.

%\afterpage{\FloatBarrier}

Our code is written entirely in \textsc{python}; to implement Kalman filters and wavelet decompositions, we have used the libraries \emph{filterpy} \citep{labbe2014} and \emph{pywavelets} \citep{lee2006}, respectively. The complete codebase and datasets used may be found online.\footnote{\url{https://github.com/astrogilda/stellar_denoising}}

\begin{figure*}
%\begin{center}
\centering
\includegraphics[width=1\linewidth]{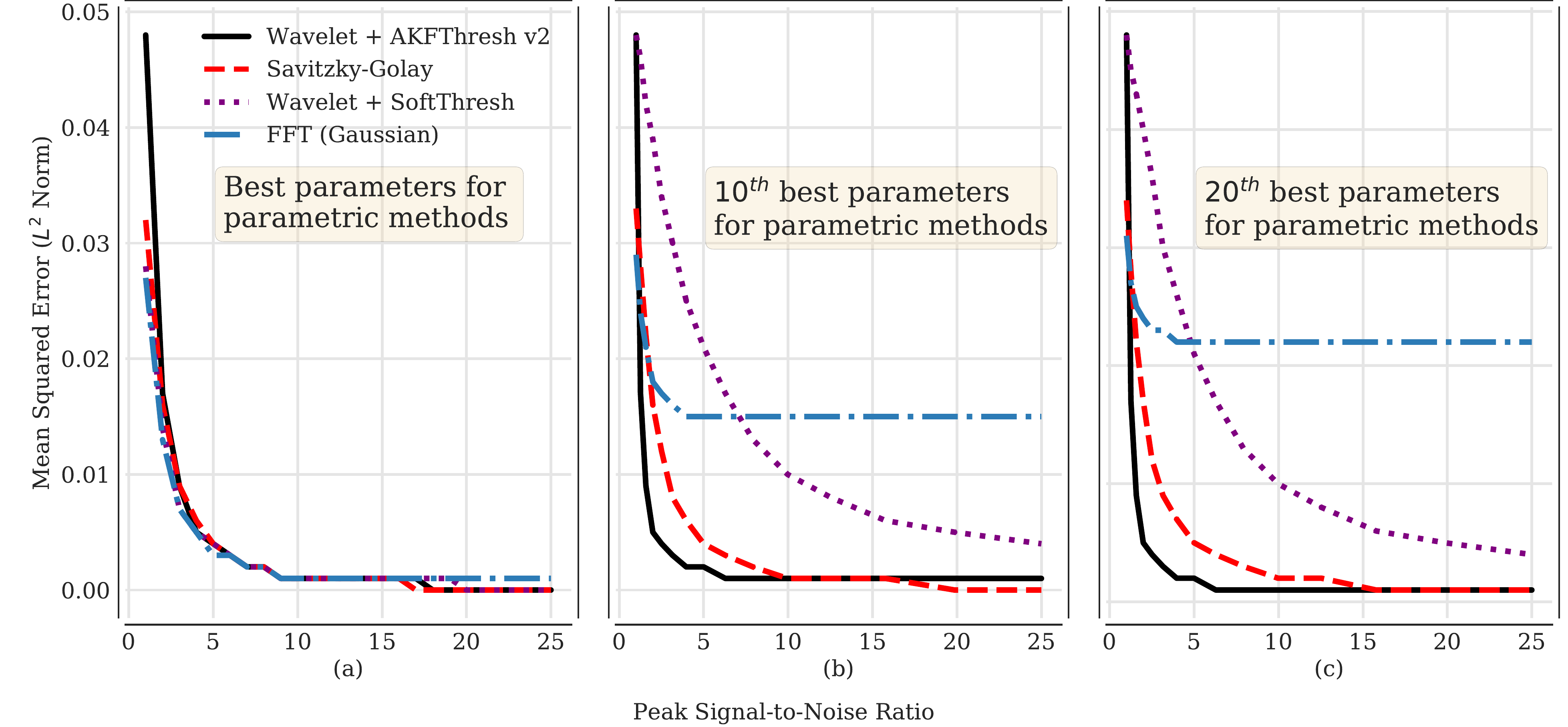}
\caption{Here we show the Mean Squared Error (MSE or $L^2$ norm) as a function of the input signal Peak Signal-to-Noise Ratio (PSNR) for the four best-performing methods. For all parametric methods, the parameters have been optimized under the same metric used for this comparison, i.e. MSE. {\it (a)} Performance with \emph{optimal} parameters for the parametric methods (Savitzky-Golay, Wavelet + SoftThresh, FFT with Gaussian kernel), {\it (b)} performance with the \emph{tenth-best} set of parameters, and {\it (c)} performance with the \emph{twentieth-best} set of parameters. In the leftmost and middle panels, our algorithm (solid black) lies essentially directly under the curve (dashed red) for Savitzky-Golay. When the parametric methods are used with their optimal parameters, they perform essentially the same as our algorithm. However, using the tenth-best parameters substantially harms their performance relative to our algorithm, with the difference from our method (solid black) getting larger with increasing PSNR for FFT (Gaussian) (solid blue), and growing and then slightly decreasing for Wavelet + SoftThresh (short-dashed purple). At tenth-best parameters, Savitzky-Golay (dashed red) remains comparable to our method, but our method performs slightly better when the twentieth-best parameters for Savitzky-Golay are used (rightmost panel).}% The optimal parameters for the parametric methods are given in Table \ref{table:tab1}.}

%Sankalp: Please update table reference above.
\label{fig:fig4}
\includegraphics[width=1\linewidth]{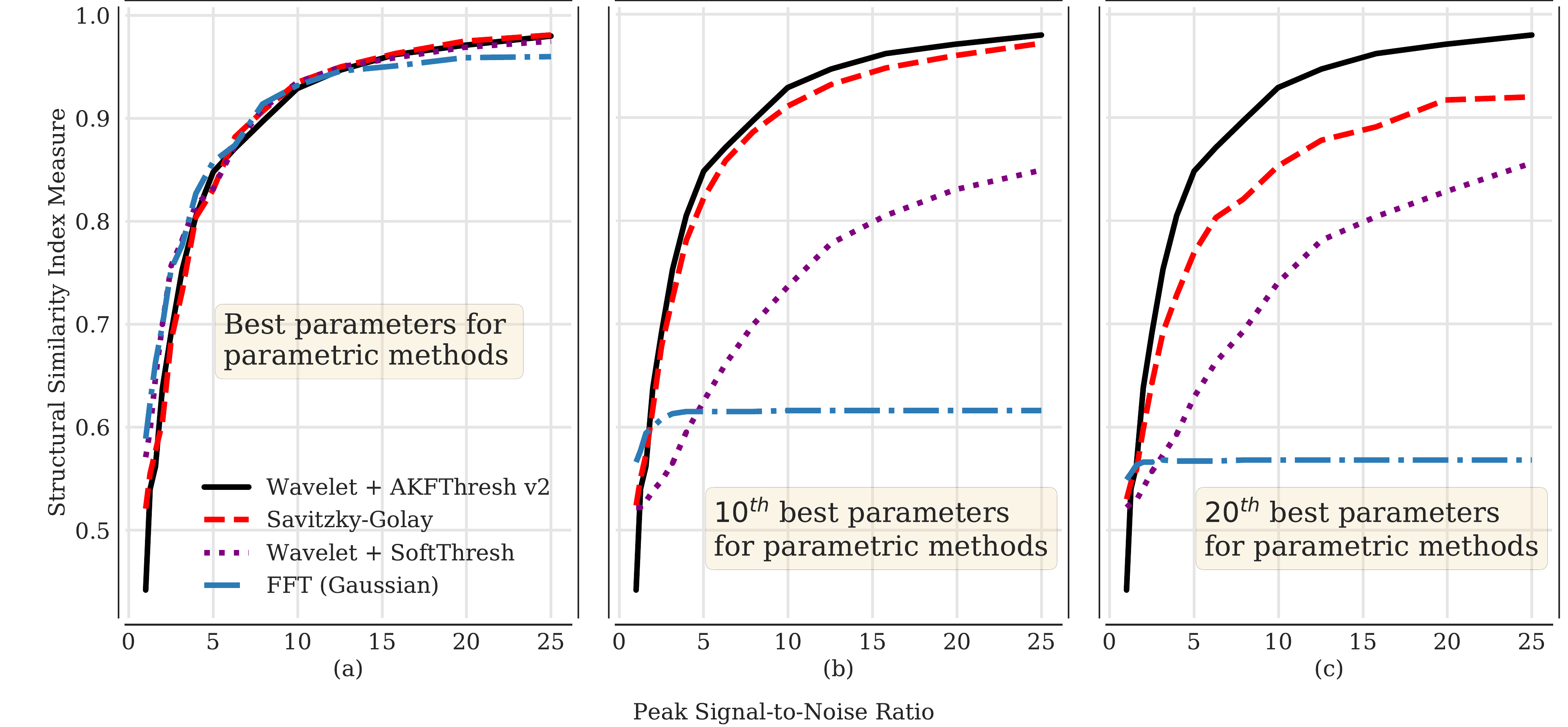}
\caption{Same as Fig. \ref{fig:fig4}, but with the Structural Similarity Index Measure as the error metric chosen both to optimize parameters and test performance for Savitzky-Golay, Wavelet + SoftThresh, and FFT with Gaussian kernel. Fig. \ref{fig:mse_vs_ssim} illustrates what SSIM vs. MSE means; higher SSIM is better. Notably, here even at tenth-best parameters our method begins to slightly outperform Savitzky-Golay, and clearly greatly outperforms the other methods. For the twentieth-best parameters, our method significantly outperforms all of the others. This observation emphasizes the importance of performing comparisons using several different error metrics.}% The optimal parameters for the parametric methods are given in Table \ref{table:tab1}.}
%Sankalp: Please update table reference above.
\label{fig:fig5}
%\end{center}
\end{figure*}

\begin{figure*}
\begin{center}
\includegraphics[width=1\linewidth]{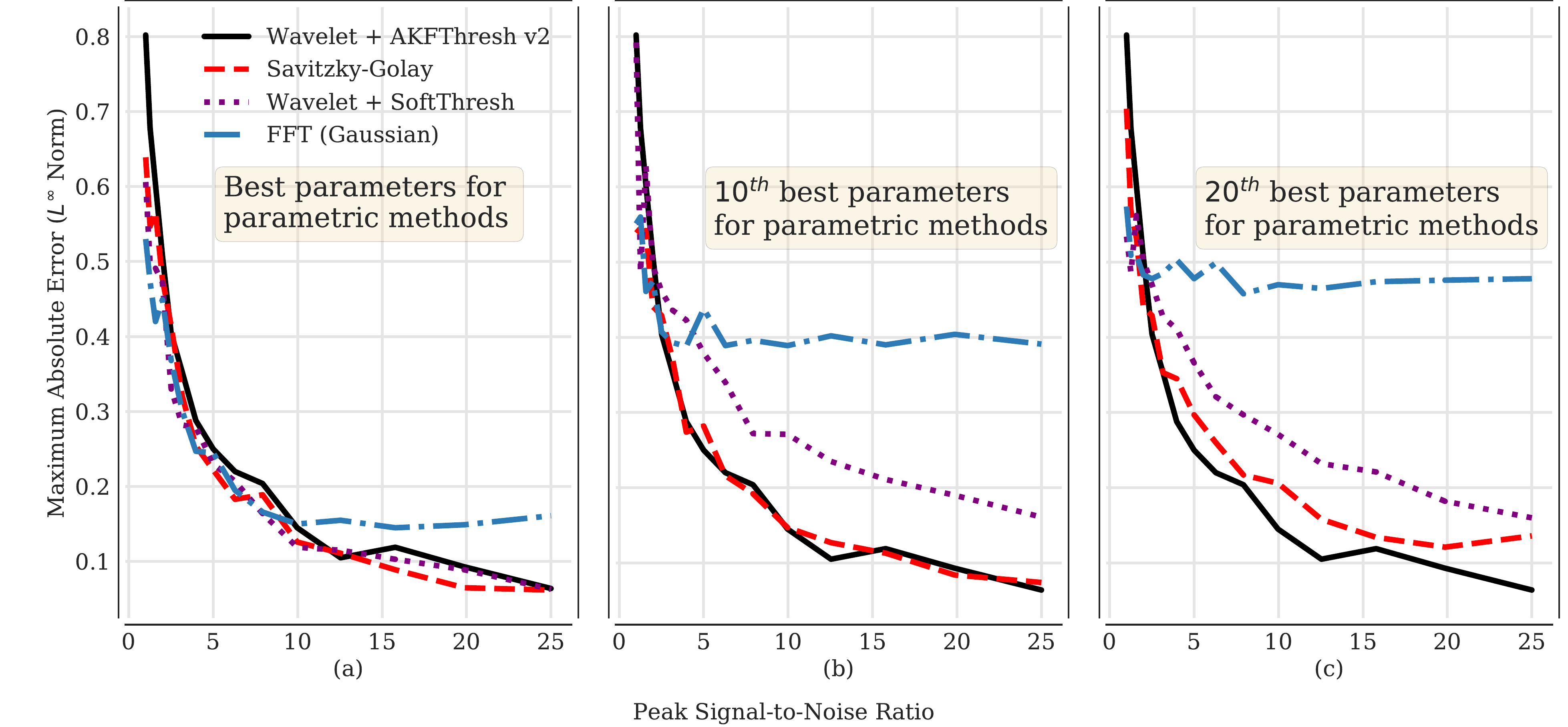}
\caption{Same as Fig. \ref{fig:fig4}, but with Maximum Absolute Error as the error metric. We again see that the methods are comparable (save for FFT (Gaussian)) when optimal parameters are used, but that our algorithm performs better than the others when sub-optimal parameters are used for them (rightmost panel, {\it c}).}
\label{fig:fig6}
\includegraphics[width=1\linewidth]{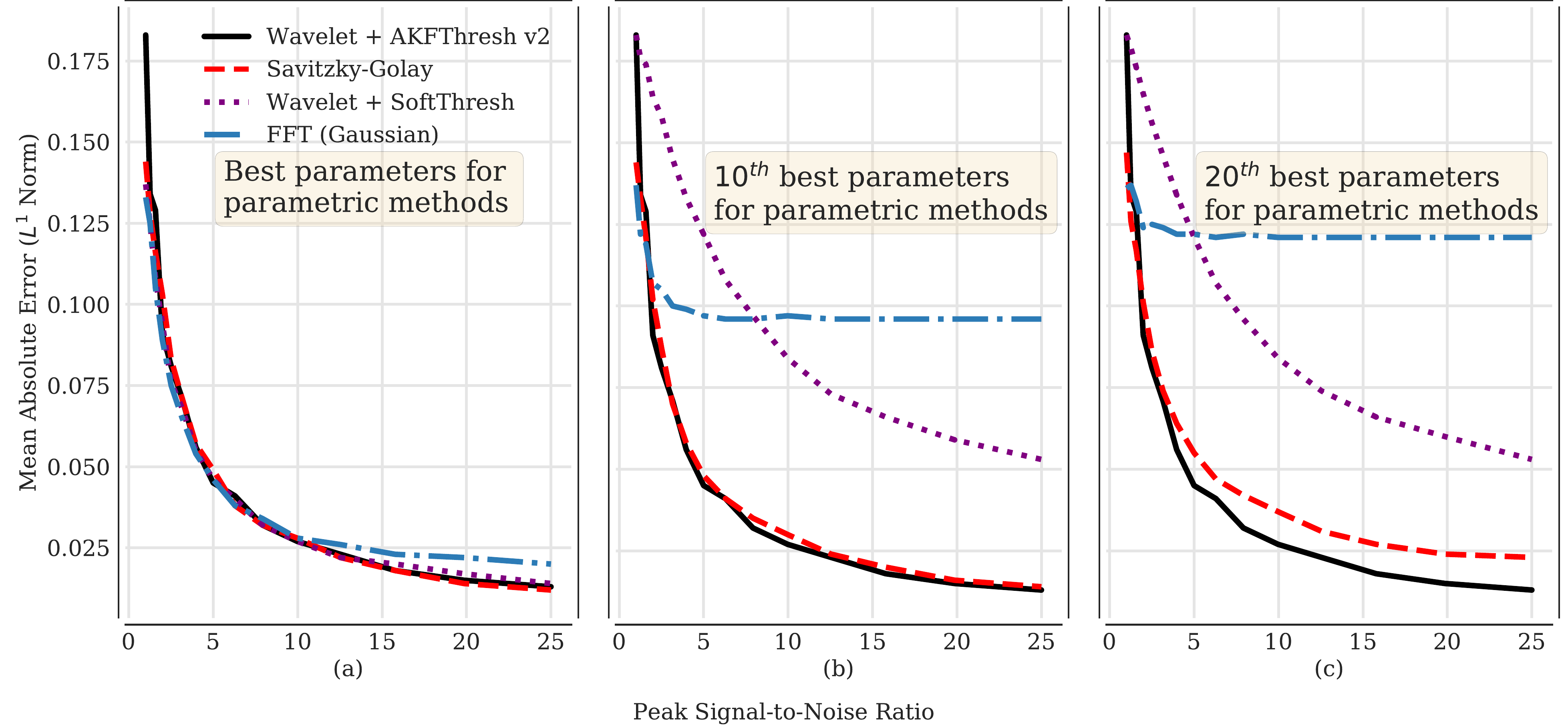}
\caption{Same as Fig. \ref{fig:fig4}, but with Mean Absolute Error as the error metric. This simply illustrates under another error metric the pattern shown by the previous Figures: our method outperforms the others when sub-optimal parameters are used in them.}
\label{fig:fig7}
\end{center}
\end{figure*}

\begin{figure*}
\centering
%\begin{center}
\includegraphics[width=.95\linewidth]{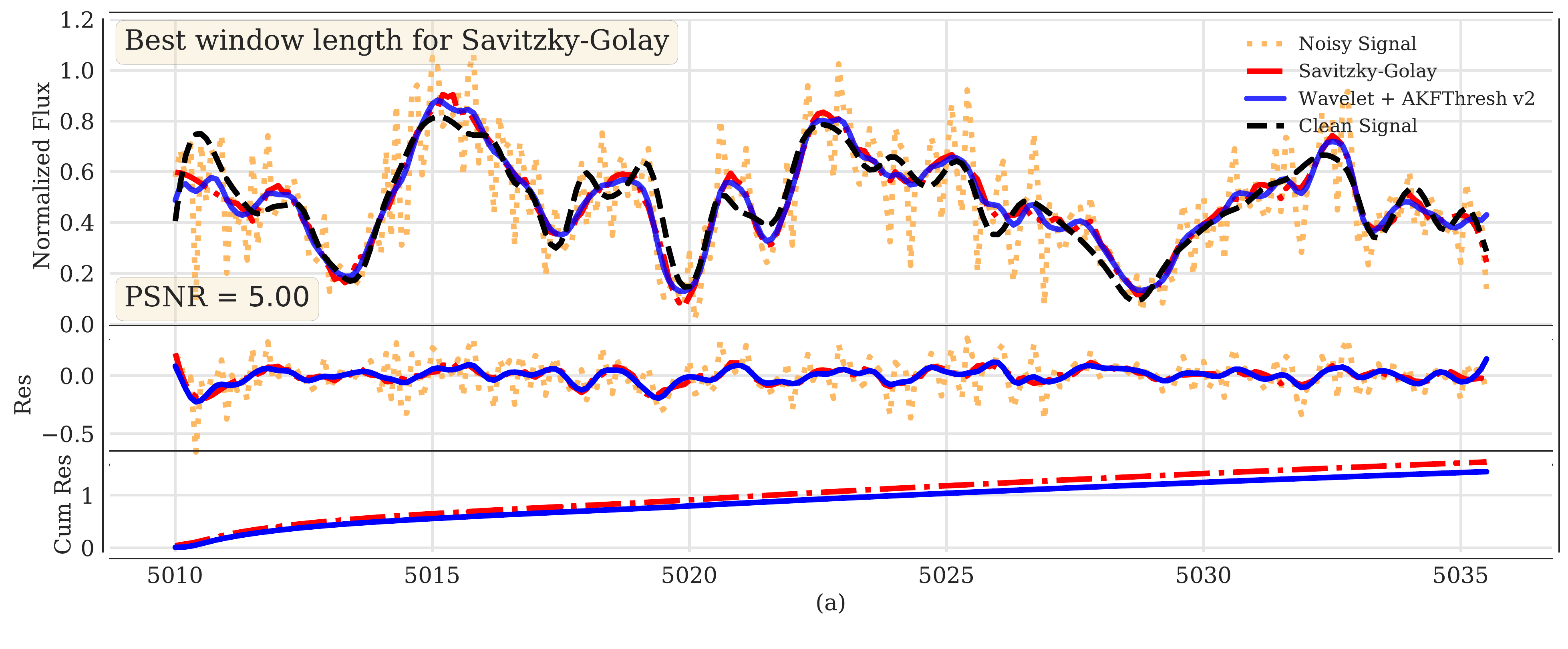}
\includegraphics[width=.95\linewidth]{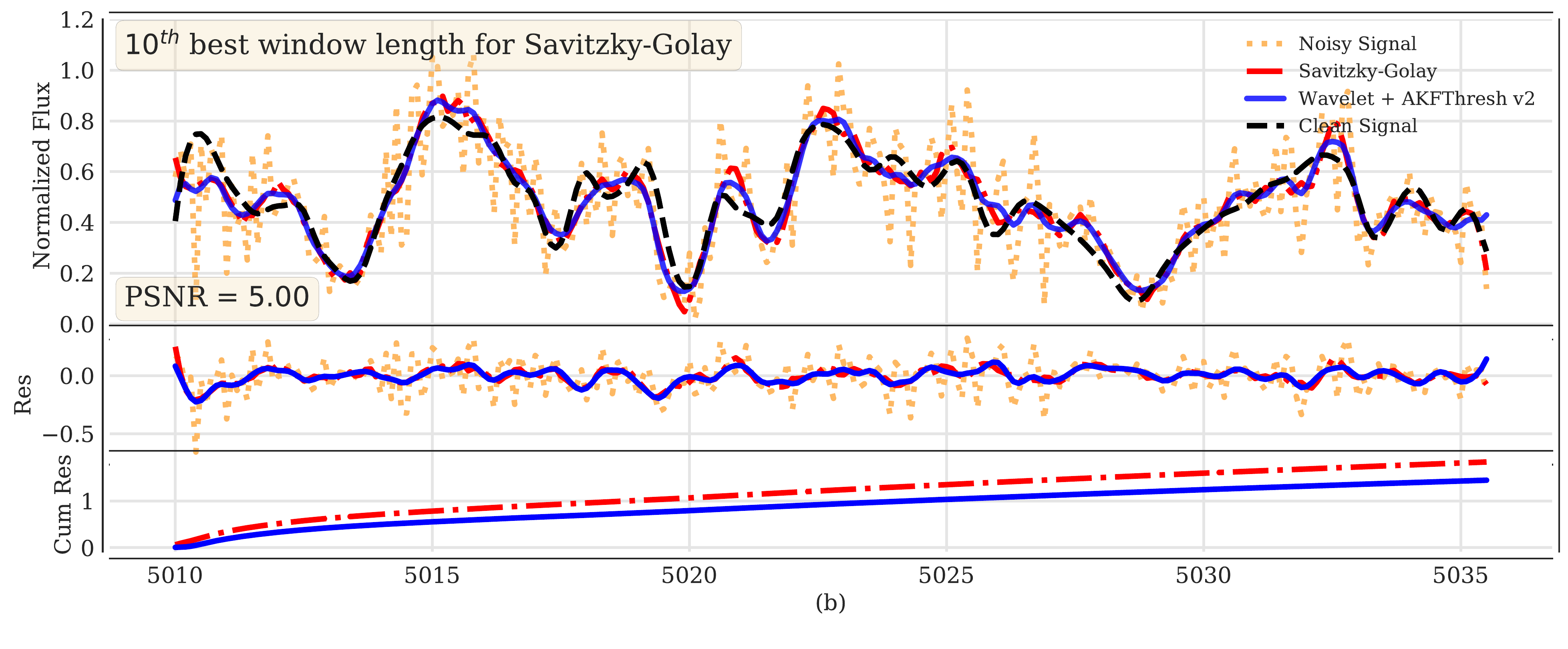}
\includegraphics[width=.95\linewidth]{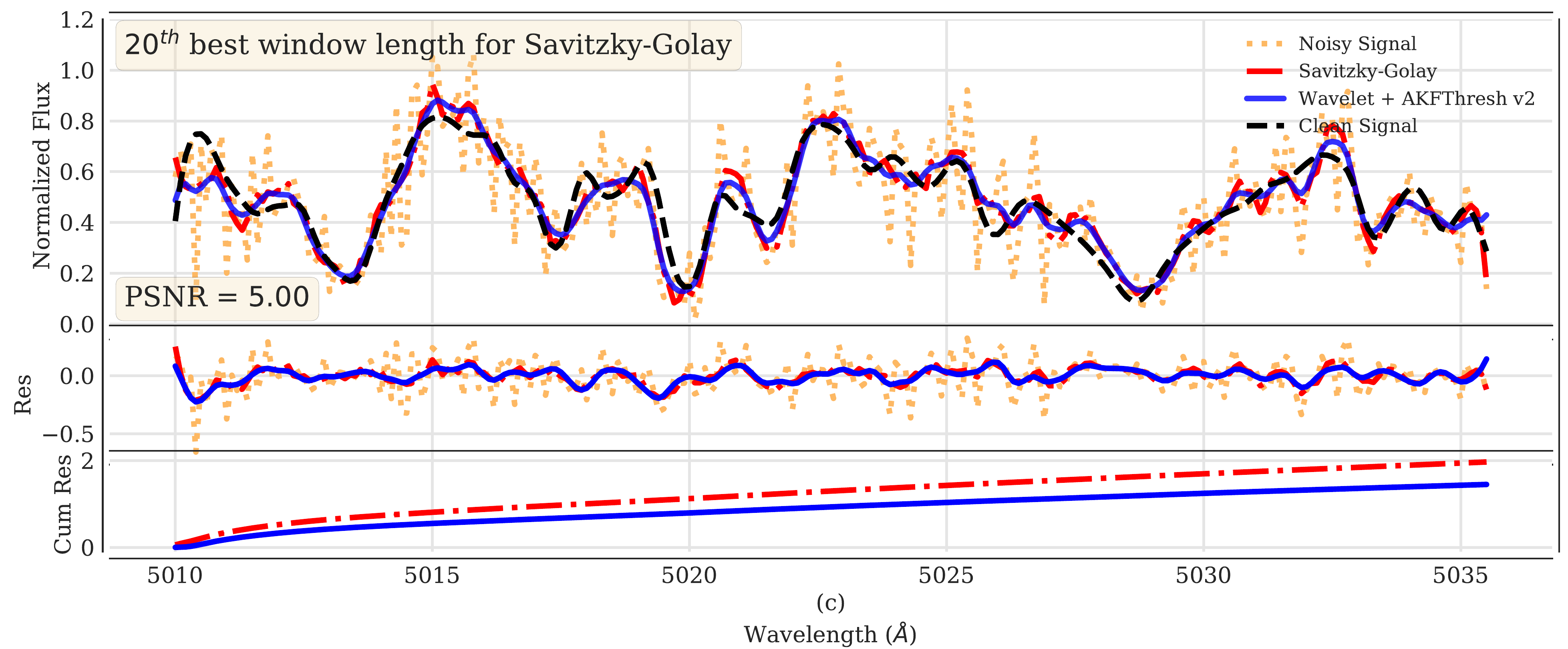}
\caption{Here we show a $25\AA$ width zoom-in, from $5010\AA-5035\AA$, on the original (clean), noisy, and denoised spectra for \emph{Star 2}. PSNR $= 5.0$ for the noisy spectrum. {\it (a)} Denoised spectra using both v2 of the proposed algorithm and Savitzky-Golay (SG) with \emph{optimal} parameters chosen by minimizing the $L^2$ norm with respect to the clean signal. {\it (b)} Same as {\it (a)} but with the {\it tenth-best} set of parameters for SG. {\it (c)} Same as {\it (a)} but with \emph{twentieth-best} set of parameters for SG. When SG has the optimal window length, it just slightly underperforms our method, as can especially be seen from the cumulative residual (summed up as one moves right in wavelength---lowermost panel in {\it (a)}). As we move to tenth and then twentieth-best SG parameters, our method outperforms it more and more, shown by the lower cumulative residual (solid blue vs. dashed red) in the lowermost panel of each sub-figure.}
\label{fig:fig10}
%\end{center}
\end{figure*}

\begin{figure*}
\begin{multicols}{2}
    \includegraphics[width=\linewidth]{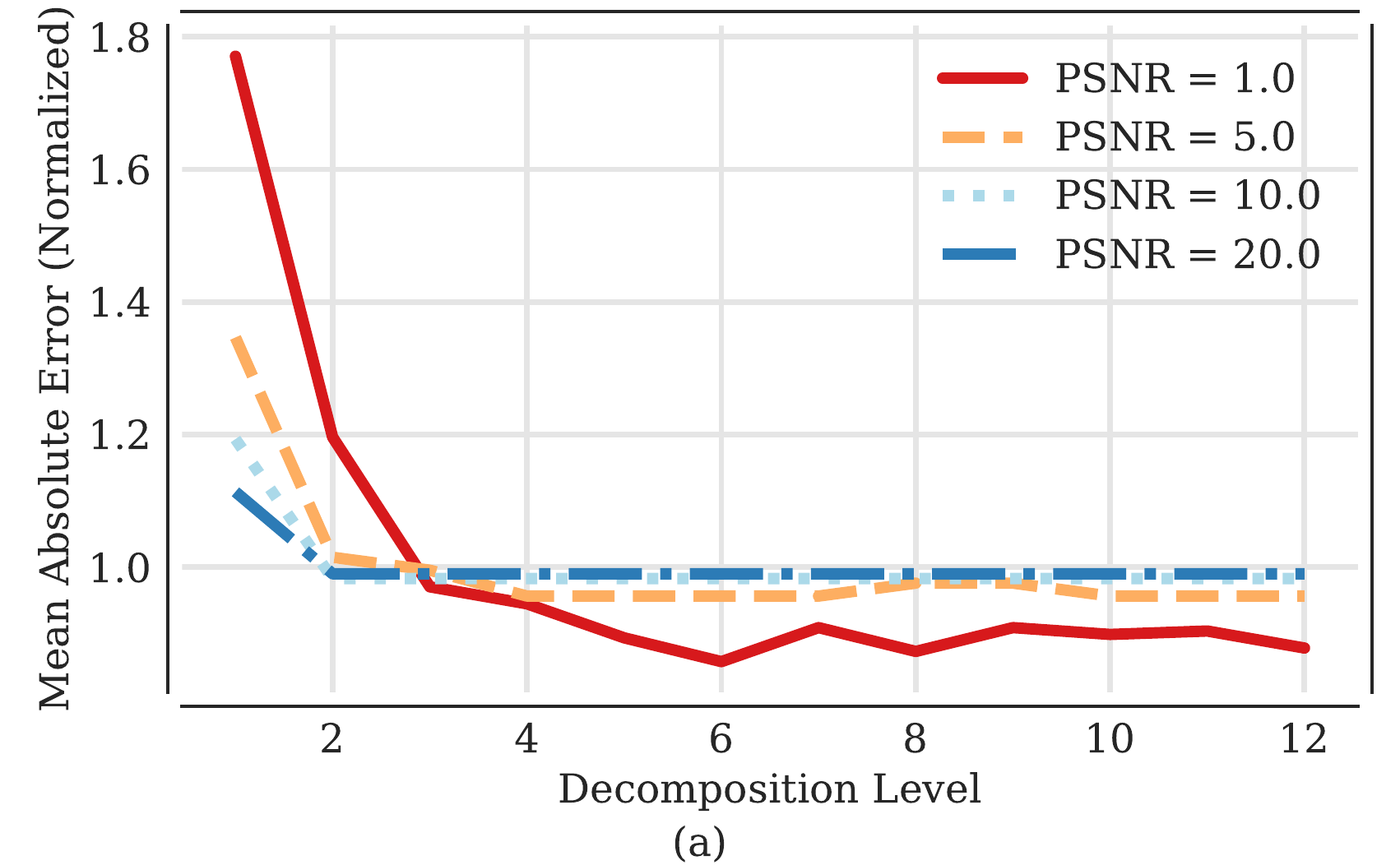}\par 
    \includegraphics[width=\linewidth]{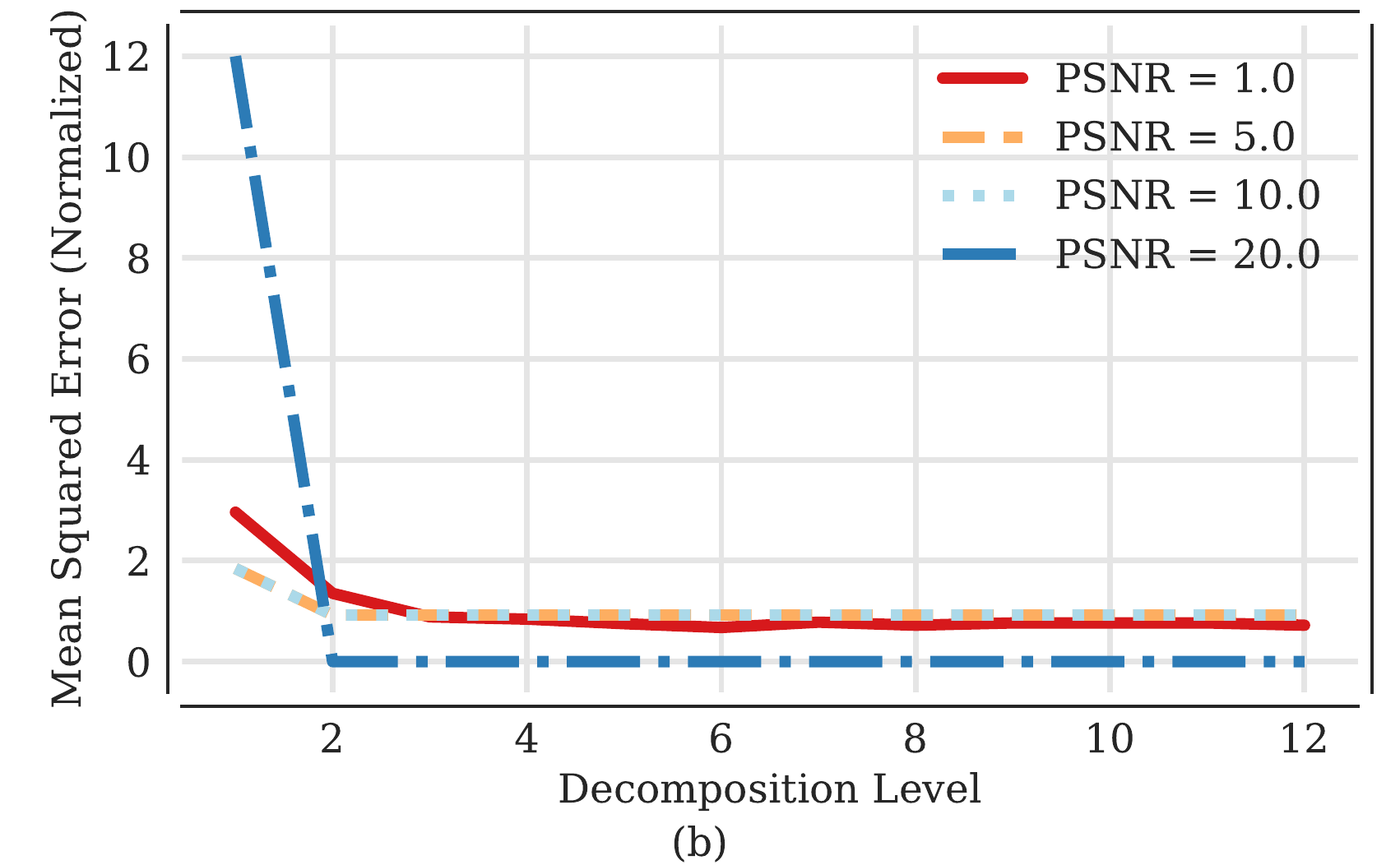}\par 
\end{multicols}

\begin{multicols}{2}
    \includegraphics[width=\linewidth]{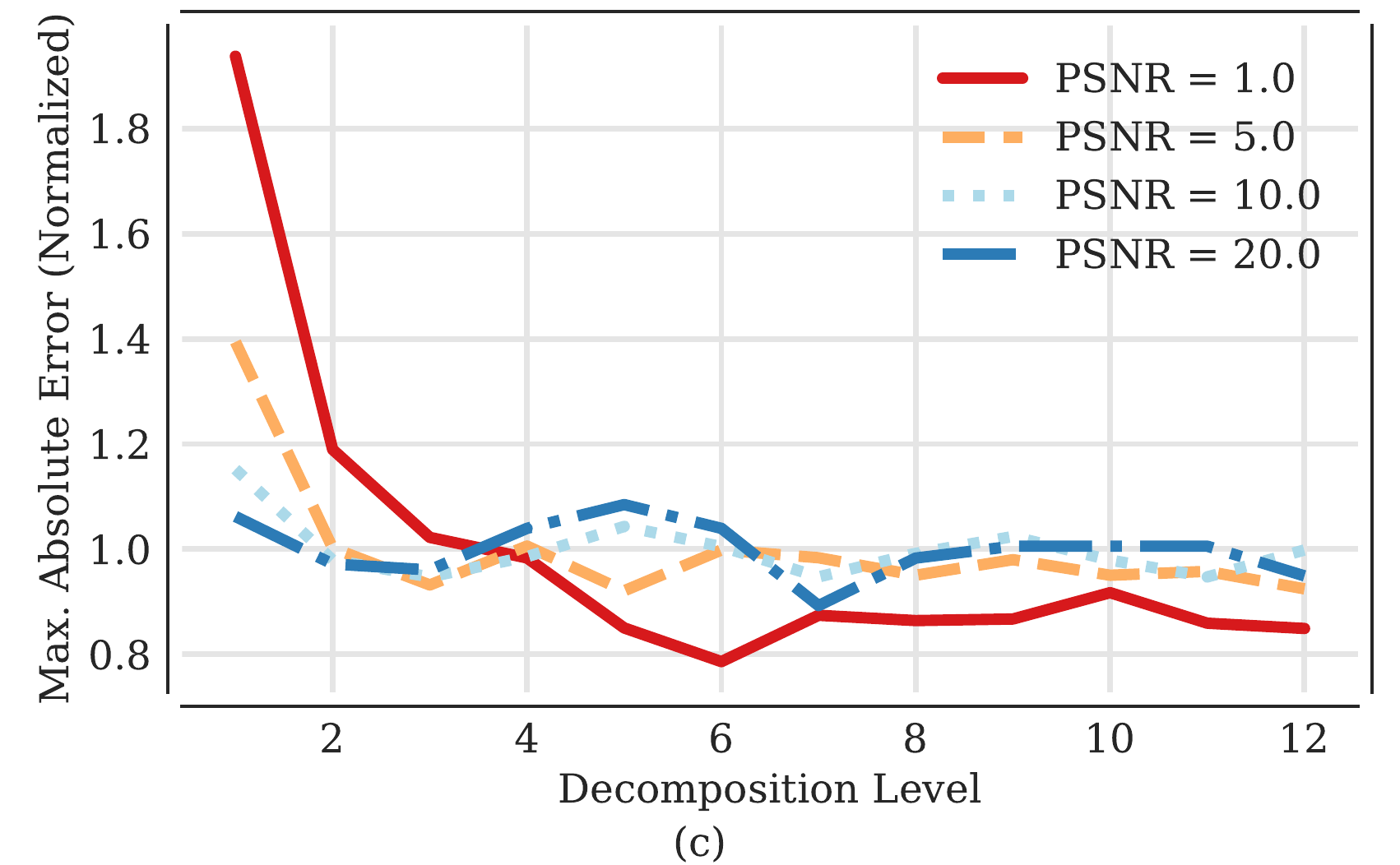}\par
    \includegraphics[width=\linewidth]{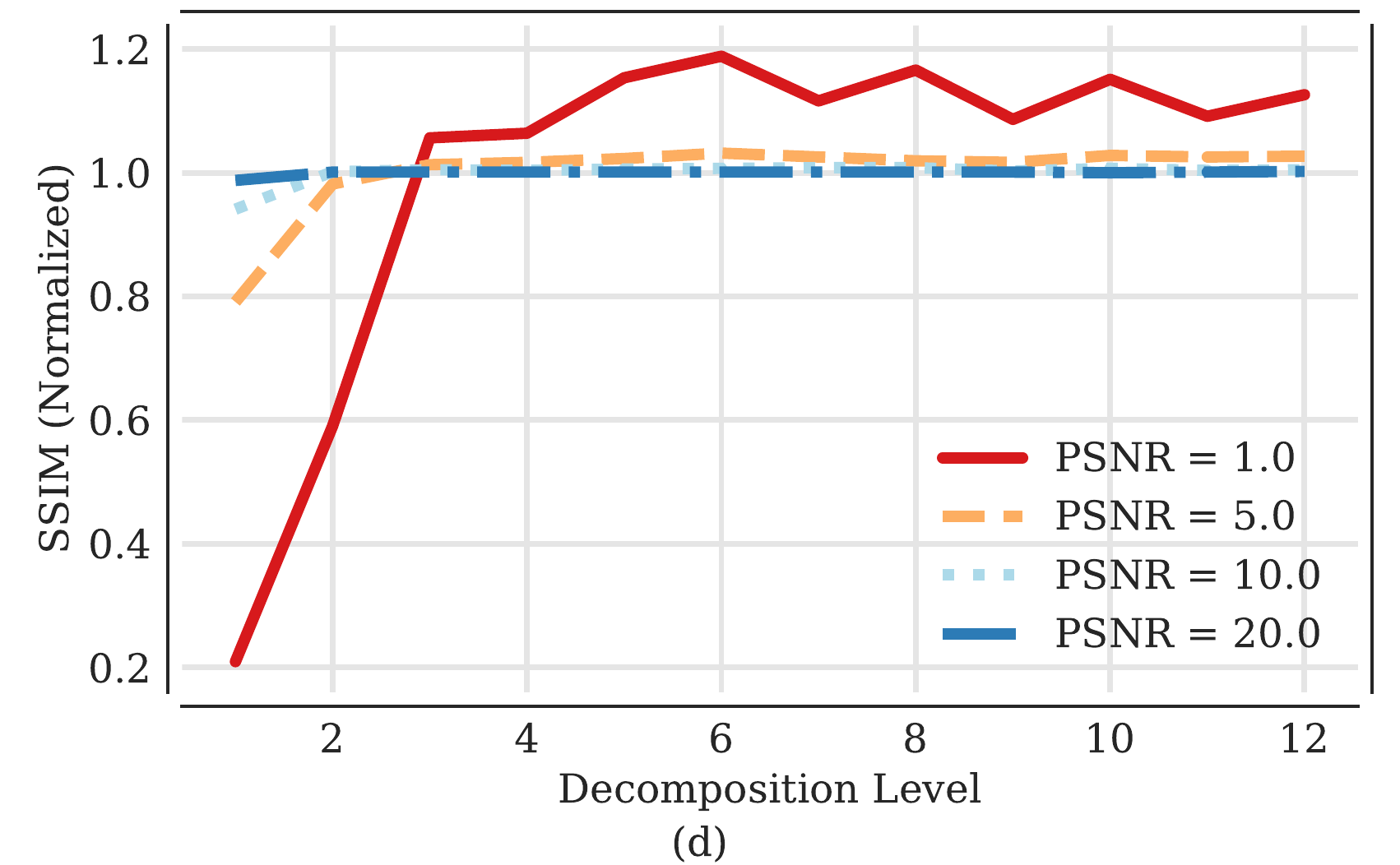}\par
\end{multicols}
\caption{Performance of Wavelet + \textsc{AKFThresh v1} as a function of the `Haar' wavelet decomposition level under different error metrics and with different PSNRs. {\it (a)} Error metric = mean absolute error. {\it (b)} Error metric = mean squared error. {\it (c)} Error metric = maximum absolute error. {\it (d)} Error metric = structural similarity index measure. At all PSNRs and for all error metrics, the denoising performance of the proposed method saturates at a wavelet decomposition level of $\approx 6$ (half the maximum possible number of decomposition levels). This justifies our choice of $6$ as the level to which to decompose the noisy input signal.}
\label{fig:fig8}
\end{figure*}

\begin{figure*}
\begin{multicols}{2}
    \includegraphics[width=.9\linewidth]{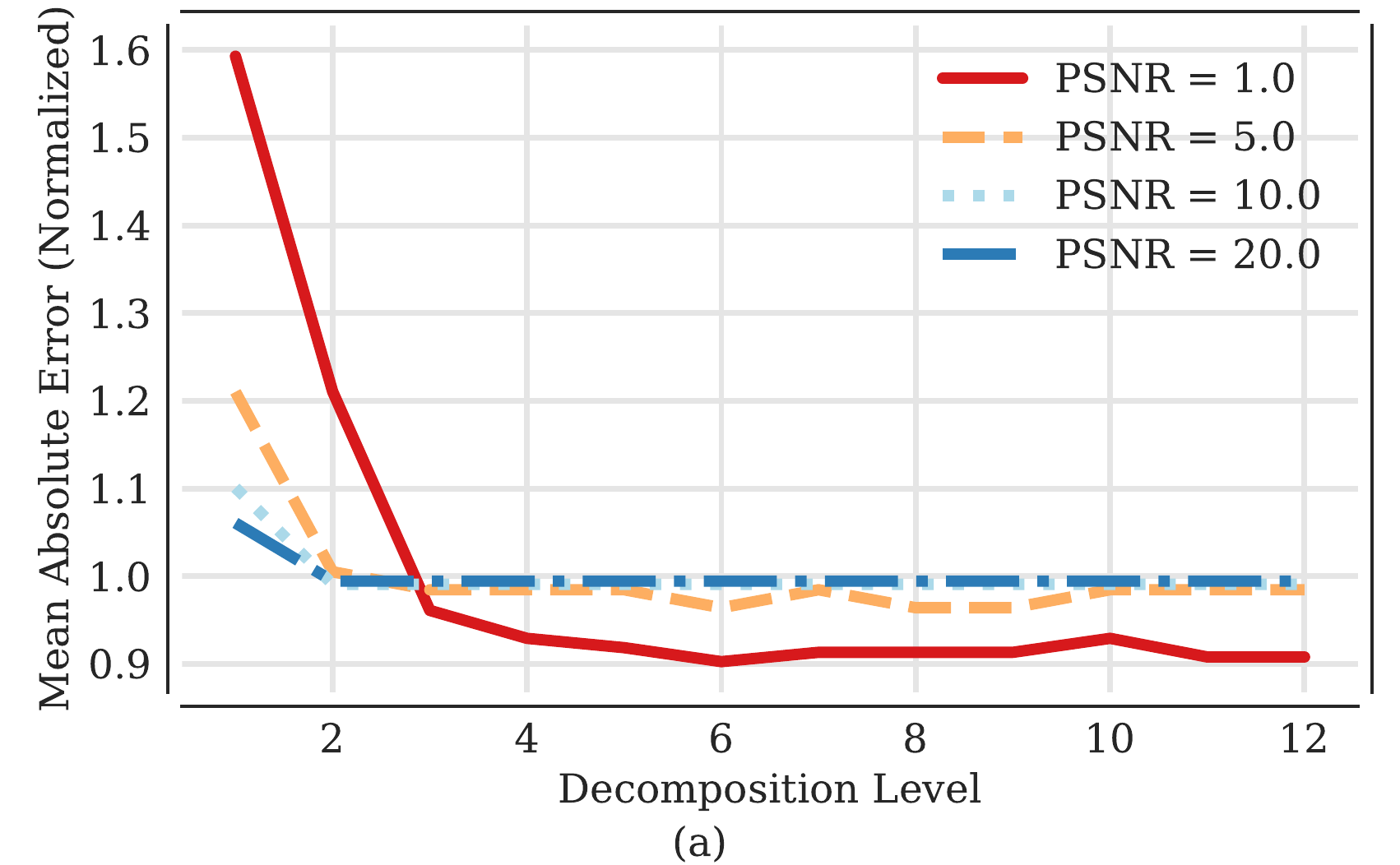}\par 
    \includegraphics[width=.9\linewidth]{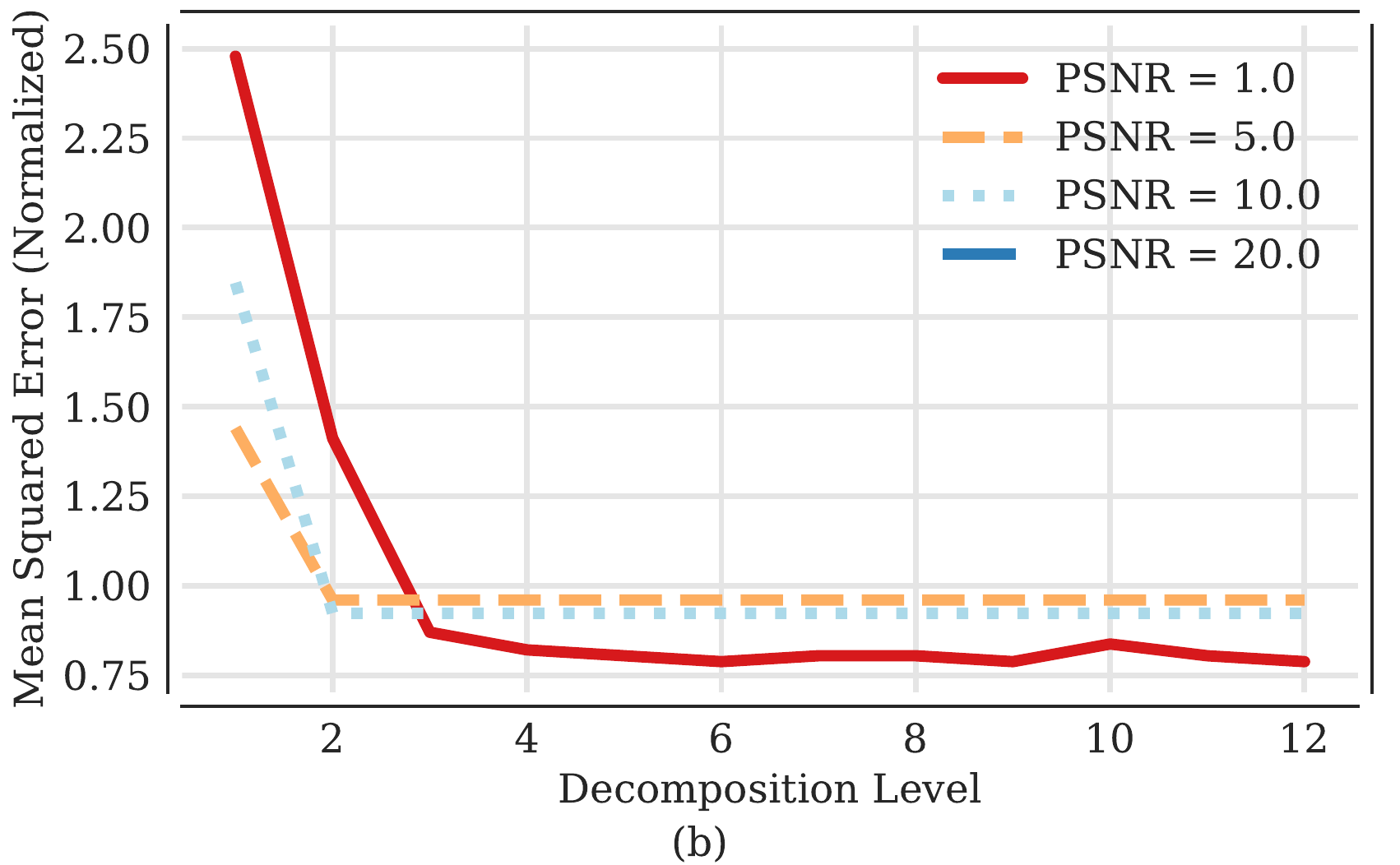}\par 
\end{multicols}
\begin{multicols}{2}
    \includegraphics[width=.9\linewidth]{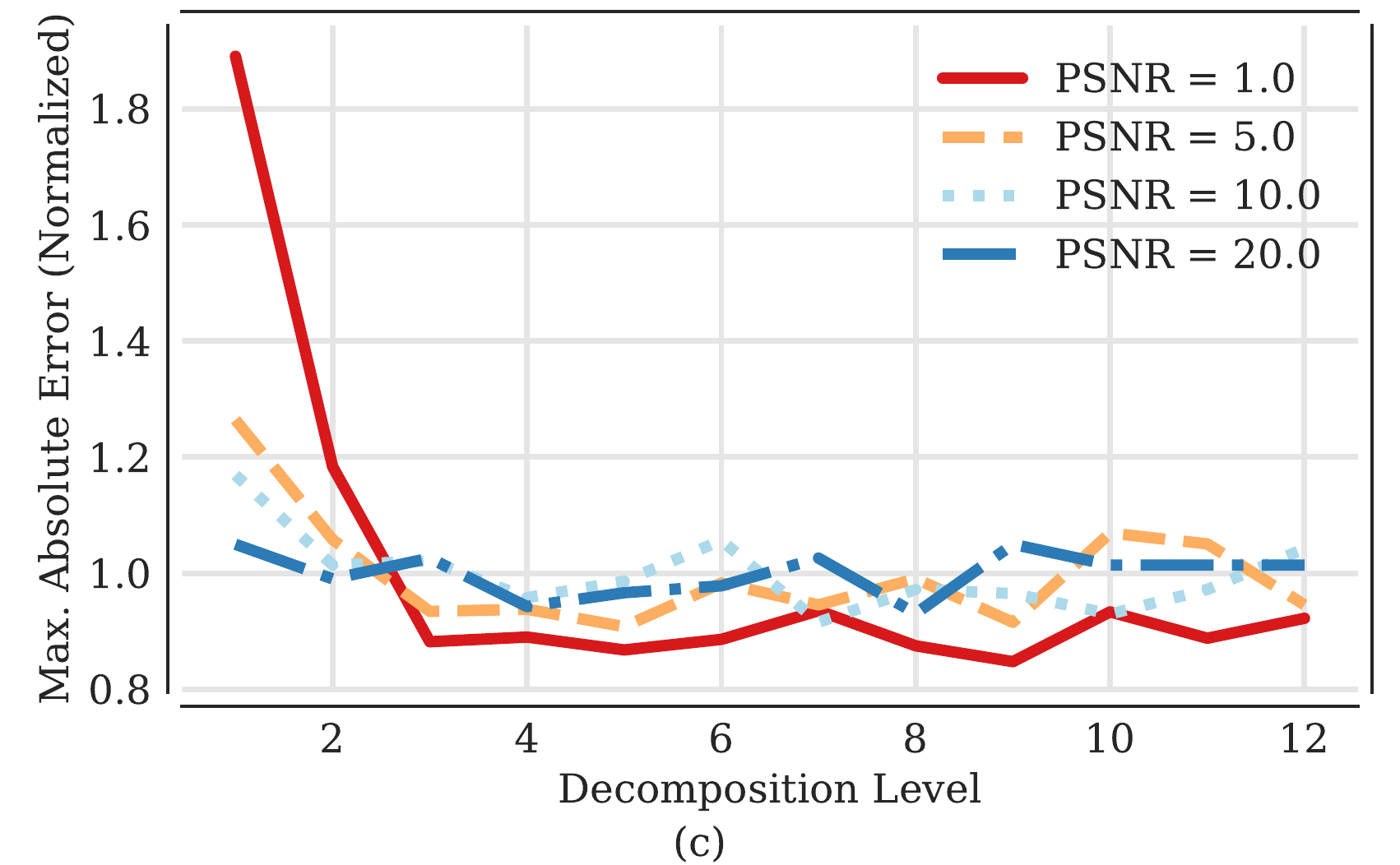}\par
    \includegraphics[width=.9\linewidth]{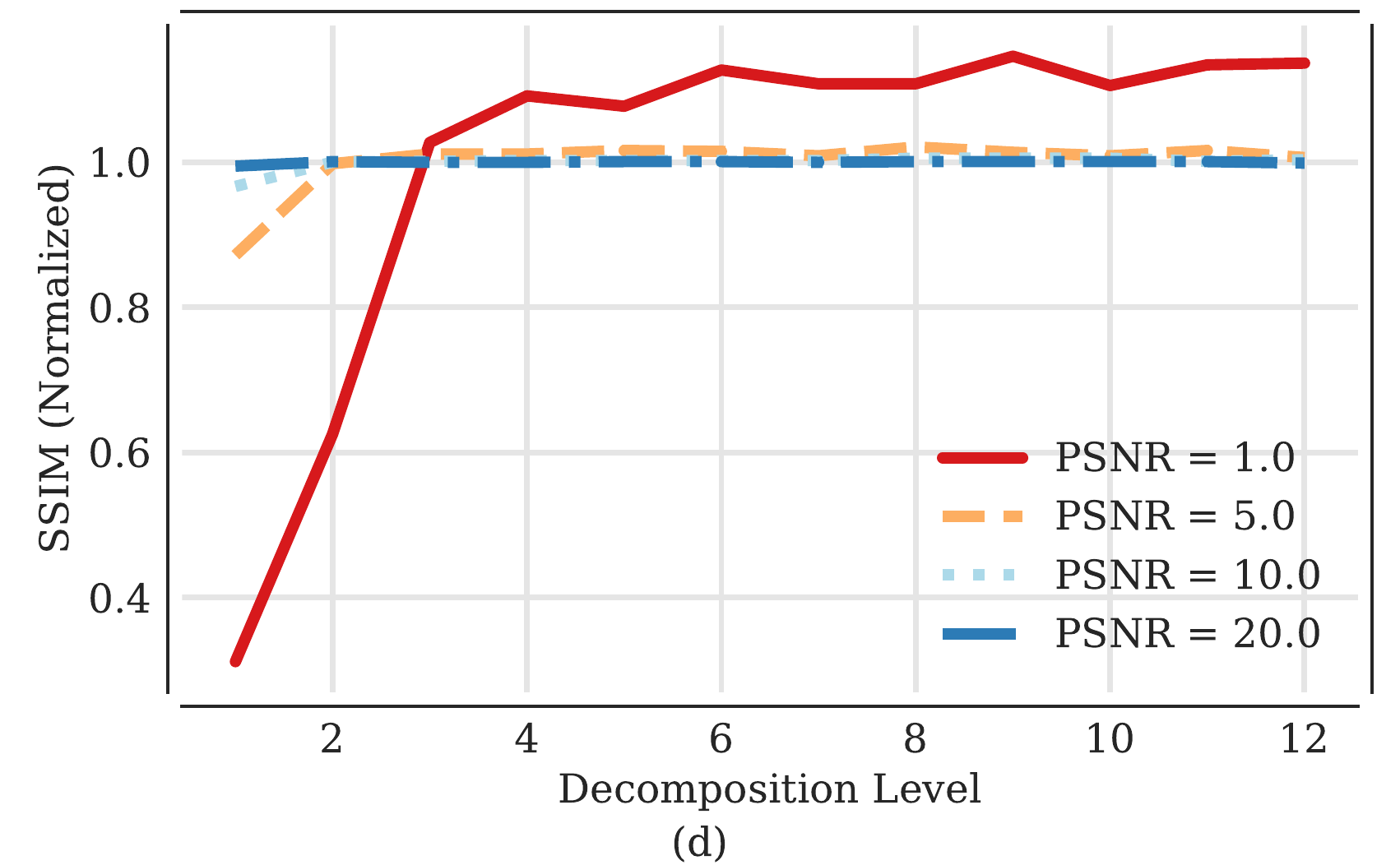}\par
\end{multicols}
\caption{Same as Fig. \ref{fig:fig8}, for Wavelet + \textsc{AKFThresh v2}, i.e. the second version of our algorithm that builds on the input of the first. We note that in the lower-righthand plot the light blue curve for PSNR = $10.0$ lies under that for PSNR = $20.0$, and in the upper righthand plot the PSNR = $20.0$ curve lies directly below the PSNR = $1.0$ curve. One can see that the performance of \textsc{v2} is generally slightly better than that of \textsc{v1} by comparing panel-by-panel to Fig. \ref{fig:fig8}. Again, after 6 decomposition levels the performance roughly saturates.}
\label{fig:fig9}

\end{figure*}

\section{Results}\label{results}
We discuss here the relative denoising performances of the methods introduced in \S\ref{akft}. In the interest of saving space, we only present results for synthetic spectra of \emph{Star 2} (T$_{\rm eff} = 4,000$K, [Fe/H] = $1.0$ dex, $\log g = 2.0$ dex, spectral type = KIII, further details in \S\ref{test_data}); the results for the other two stars are similar.

Tables \ref{table:tab11} through \ref{table:tab33}\footnote{\url{https://github.com/astrogilda/stellar_denoising}} present the values of the different comparative metrics, at three different noise levels, for the \emph{optimal} and \emph{sub-optimal} parameter cases, at a resolution of $5,000$ (again, the results are qualitatively similar at the other two resolutions of $2,000$ and $10,000$). Each table shows the mean value and the standard deviation of the results for the $100$ different runs of the respective method. %The last column in each table displays the run times for the various algorithms on a modern laptop with an $8^{th}$ generation Intel i7 processor and 8Gb of RAM.
%Sankalp: I liked having the runtimes in! Let's restore them before journal submission.
The best and the worst values for each column are highlighted---the best in \textbf{boldface} and the worst in {\it italics}. In addition, the median parameter values used for the parametric denoising methods have been specified.

In Figs. \ref{fig:fig4}, \ref{fig:fig5}, \ref{fig:fig6} and \ref{fig:fig7}, we present the values of the four performance metrics for the best four denoising algorithms---FFT with Gaussian kernel, Savitzky-Golay, Wavelet + Soft Thresholding, and Wavelet + \textsc{AKFThresh v2}---as a function of input signal PSNR and quality of parameters for the different denoising methods. The highest quality parameter is that which gives the lowest $L^2$ for a given method, the second-highest quality parameter that which gives the second lowest $L^2$, and so on. As the standard deviations are relatively small, they have been omitted for clarity.

In Fig. \ref{fig:fig10}, Fig. \ref{fig:fig11}, and Fig. \ref{fig:fig12} we compare the denoising performance of the best version of the proposed algorithm (\textsc{v2}) with that of Fourier transform with \emph{optimal} and \emph{sub-optimal} parameters. We have only shown a small section of the input spectrum to better emphasize the differences between the different spectra. At PSNRs $\geq 5.0$, our algorithm is able to extract the peaks and valleys of the original input spectrum from the noisy spectrum much better than the Fourier transform. Its superiority in filtering out the Poisson noise is especially highlighted when the width of the Gaussian kernel for the FFT is not optimal for the specific noisy spectrum. This is indeed likely to be the case when the user is responsible for picking it without access to the original spectrum.
%For each of these figures, parts (a) denote the outcomes when the parameters for the compared methods are optimized for best performance (i.e., lowest $L^2$ norm), while parts (b) denote the outcomes when the parameters resulting in second-lowest $L^2$ scores are chosen. Figures 7 through 8  repeat this analysis, but with the Structural Similarity Index Measure ($SSIM$) plotted on the y-axis.

Finally, in Figs. \ref{fig:fig8} and \ref{fig:fig9}, we show the performance of both versions of our denoising methods as a function of the decomposition level. Performance as measured by all four error metrics more or less saturates once decomposition exceeds the $6^{\rm th}$ layer. This saturation justifies our decision, and recommendation, to decompose the input noisy signal to half the maximum possible number of decomposition levels. In this work, the maximum number of levels possible is $12$ because we had spectra of $4096 = 2^{12}$ wavelength bins).

Based on these results, we can draw several conclusions:
\begin{enumerate}
    \item At almost all noise levels and all choices of parameters, exponential smoothing has the worst performance.
    \item DWT with both soft and hard thresholding gives excellent results across all four metrics for the optimal choice of parameters (i.e. maximal decomposition level, thresholding), but the performance quickly deteriorates for an incorrect choice of parameters. We emphasize that one does not know the correct parameters \emph{a priori}, so this issue is a significant one.
    \item At extremely low PSNRs ($<5$), both versions of our algorithm provide poor denoising results, even when compared to the performance of other methods with sub-optimal parameters.
    \item At PSNRs $>5$, both versions of our algorithm start outperforming other methods, especially when sub-optimal parameters have been chosen for the other methods. This scenario is the more likely one in practice since one does not always have a reference spectrum with which to work. Indeed, calibrating a parametric denoising method on synthetic spectra before applying to real data comes with its own set of issues, primarily interpolation errors when creating the most closely-matched synthetic spectrum, as well as errors originating from instrument peculiarities' not being captured in the synthetic spectra.
    \item Savitzky--Golay is the only algorithm that compares favorably to the algorithm proposed in this paper with respect to the four error metrics we explore (Figs. \ref{fig:fig4}--\ref{fig:fig7}). However, we note that during parameter selection for the commonly used parametric methods including Savitzky-Golay, we explored a rather wide range of parameters (e.g. for the choice of the window-length `w', see footnotes for Tables \ref{table:tab11} through \ref{table:tab33}). 
    %Sankalp: please update table reference. Also check whether figure references above are correct!
    This was computationally practical for a small number of spectral types (we considered just three different types of star in this work). However for a large survey dataset with millions to tens of millions of objects (e.g. DESI will have $\sim 10^7$ stellar spectra), with many different possible spectral types, such an exploration would likely be computationally demanding. In particular, one might, for each expected spectral type, optimize the parameters using synthetic spectra of that type. But then one would still need to build in a step where, for a given data spectrum, the correct type was automatically identified to allow the optimal parameters for that particular type for a given algorithm to be used. Having a method that simply does not require optimal parameter selection at all seems more efficient.
\end{enumerate}

\section{Conclusions and Future Directions}\label{conclusions}
In this work, we have proposed a new method for denoising 1D stellar spectra. Our proposed method involves two stages---a discrete wavelet transform, followed by adaptive Kalman-filter-based thresholding. By using the `Haar' wavelet basis, the only free parameter for the user is the number of levels of decomposition for the wavelet transform. By setting this to half the maximum possible value (where the improvement of adding more levels essentially saturates), the number of free parameters can in fact be reduced to zero. Our proposed method has been validated on several types of synthetic PHOENIX spectra with Poisson noise injected. Our tests show that the method can improve denoising better than conventionally employed tools (up to $3\times$ if the conventional methods have even slightly sub-optimal parameters; for instance as shown in Figs. \ref{fig:fig4}--\ref{fig:fig7}). The method performs well all the way down to a PSNR of roughly 5. 

Our method may also be useful denoising galaxy and quasar spectra, which often have low SNRs ($<20$). There are many large current and upcoming such datasets that offer fruitful use-cases: the Sloan Digital Sky Survey (SDSS), Gaia, the Large Synoptic Survey Telescope (LSST), the Public ESO Spectroscopic Survey for Transient Objects (PESSTO), the `deep VIMOS survey of the CANDELS UDS and CDFS fields' (VANDELS), the Large Early Galaxy Astrophysics Census survey (LEGA-C), and the Dark Energy Spectroscopic Instrument survey (DESI). DESI in particular will have low SNR, low-resolution spectra of roughly 30 million galaxies and quasars, and denoising these might aid both in redshift fitting for large-scale structure clustering analyses as well as in ancillary science. Denoising spectra of the Lyman-$\alpha$ forest as measured using quasars as ``skewers'' might also enhance the science possible e.g. regarding warm dark matter constraints (for instance, \citealt{Viel2013}, \citealt{Irsic2017}) or intergalactic medium physics (e.g. \citealt{gurvich2017}). 

We note that although our proposed algorithm has been tested and validated on stellar absorption spectra, we believe it would perform just as well on emission spectra, which can be thought of as `flipped' absorption spectra. In future work, we plan to evaluate the performance of \textsc{AKFThresh v2} on spectra from the public data releases of the surveys listed above.%All in all, the proposed AKFThresh-based wavelet filter has been shown to outperform standard denoising techniques while also requiring fewer parameters to be arbitrarily (and possibly wrongly) tuned.

The current work is just a first step in developing better denoising algorithms based on sophisticated statistical methods, drawing on advances in the fields of digital signal processing \citep{oktar2016speech}, autonomous vehicle control \citep{kasper2008sensor}, and financial timeseries analysis \citep{renaud2005wavelet}, among others. There is consequently significant scope for improvement and development. For instance, we are currently updating the linear Kalman filter employed here to the Unscented Kalman filter capable of modelling non-linear state-space systems \citep{wan2000unscented, deng2013adaptive}. We are also exploring the impact of developing and using a custom mother wavelet instead of the `Haar' wavelet. Another possible line of research would be to apply Kalman filtering to reconstructed decomposition levels, estimating the noise and process covariances via an ARMA process \citep{huang2015auto}. One might also investigate deviations from the assumption of Gaussianity for the process or measurement noise, or of the assumption of zero covariance between them \citep{liu2016bayesian}. These ideas might enable denoising down to lower PSNRs, $<5$.

In addition to wavelet shrinkage as demonstrated here, one might also explore the application of Kalman filtering in conjunction with other DWT-based methods like wavelet packet thresholding \citep{oktar2016speech, zhao2017adaptive}, wavelet coefficient modeling \citep{chen2003multiwavelets, jiang2010improved, kumar2012improved}, and the wavelet transform modulus maxima (WTMM) method \citep{liu2007new}. Finally, we see no reason that this method cannot be extended to denoising astronomical time series, though this would require analysis of the data's properties to develop an appropriate basis wavelet.

\section*{Acknowledgements}
We thank Stephen Eikenberry and Paul Torrey for productive discussions, and Peter Barnes and Kai Polsterer for a careful review of the original draft. We particularly acknowledge Jian Ge for his kind guidance and mentoring over the past several years, and many helpful insights on making this paper more applicable to current science needs.

%\raggedright
\bibliography{mnras_guide.bib}

%\clearpage

\appendix{}
\section{Derivation of the Linear Kalman Filter}\label{kalman_derivation}

Here we present an intuitive derivation of the linear Kalman filter used in this work. We have presented the derivation in terms of filtering a temporal signal, as this is the traditional application of a Kalman filter. However we emphasize that in this work the Kalman filter has been used to denoise spectral data in wavelength domain, leading to applying it forwards and backwards in wavelength (as discussed in \S\ref{akft}). We have also used different, easier--to--understand symbols than in the main text above; the latter have been used to match the practice in signal processing literature. The transformations between the symbols used in the main text above and those used in the derivation here are specified in Table \ref{tab:conversion}.

% \usepackage{booktabs}
\begin{table}
\caption{Conversion between variables of the Kalman filter used in Appendix \ref {kalman_derivation} and in \S \ref{kalman} in the main body of this paper.}
\label{tab:conversion}
\begin{tabularx}{\columnwidth}{CC}%{tabular}{@{}cc@{}}
\toprule
Appendix                        & Main Text                \\ \midrule
${\bf t}_{i}$                    & ${\bf x}_{i}$           \\
${\bf n}_{i}^{t}$                & ${\bf w}_{i}$           \\
$\hat{{\bf E}}_{i}{\bf t}_{i-1}$ & ${\bf H}_{i}$           \\
${\bf m}_{i}$                    & ${\bf z}_{i}$           \\
${\bf n}_{i}^{m}$                & ${\bf v}_{i}$           \\
$\hat{{\bf M}}_{i}{\bf t}_{i-1}$ & ${\bf M}_{i}$           \\
${\bf t}_{i}^{{\rm pred}}$       & $\hat{{\bf x}}_{i}^{-}$ \\
${\bf T}_{i}^{{\rm pred}}$       & ${\bf P}_{i}^{-}$       \\
${\bf T}_{i}^{{\rm upd}}$        & ${\bf P}_{i}$           \\
${\bf N}_{i}^{t}$                & ${\bf Q}_{i}$           \\
${\bf t}_{i}^{{\rm upd}}$        & $\hat{{\bf x}}_{i}$     \\
${\bf W}_{i}$                    & ${\bf K}_{i}$ \\ \bottomrule        
\end{tabularx}
\end{table}

Kalman filtering answers the question of how one should optimally update predictions in the presence of noisy measurements. The prediction is smooth but likely, eventually (in a time series) to become biased; the measurements are, one hopes, unbiased (noise averages to zero), but noisy. How does one employ both?

Let us denote the true signal (also called the state vector) at timestep $i$ by ${\bf t}_{i}$ and the measured signal by ${\bf m}_{i}$. These are vectors, denoted by boldface. We denote the intrinsic noise, i.e. describing stochasticity of the system itself, independent of measurement, as ${\bf n}_{i}^{t},$ superscript $t$ for ``true.'' We denote the measurement noise (associated with the process of measurement) ${\bf n}_{i}^{m},$ superscript $m$ for ``measurement.'' We also define an evolution matrix ${\bf \hat{E}}_{i}$
that takes the true signal from time $i$ forward to time $i+1$. The hat represents that this matrix is an operator, and operates on a state.

We also define a measurement matrix $\hat{{\bf M}}_{i}$, where the hat denotes that it is an operator (and is useful in distinguishing it later from another matrix ${\bf M}$ for the covariance matrix of the measurement). We also note that the hat implies the operator may not commute. However, that is not problematic---in general matrix multiplication does not commute, so we are not implying an additional property beyond what should already hold.

Finally, we make the following assumptions (the derivation and operation of the Kalman filter when these assumptions do not hold is beyond the scope of this paper; interested readers are referred to the references in \S\ref{akft}):
\begin{itemize}
  \item ${\bf n}_{i}^{m}$ is 0-mean white noise: $E\left({\bf n}_{i}^{m}\right) = 0$, ${\rm cov}\left({\bf n}_{i}^{m}, {\bf n}_{j}^{m}\right)={\bf N}_{i}^{m}\delta_{ij}^{K}$
  \item ${\bf n}_{i}^{t}$ is 0-mean white noise: $E\left({\bf n}_{i}^{t}\right) = 0$, ${\rm cov}\left({\bf n}_{i}^{t}, {\bf n}_{j}^{t}\right)={\bf N}_{i}^{t}\delta_{ij}^{K}$
  \item intrinsic noise and measurement noise are uncorrelated: ${\rm cov}\left({\bf n}_{i}^{t}, {\bf n}_{i}^{m}\right)=0$
  \item true signal at timestep $i=0$ (denoted by ${\bf t}_{0}$) is uncorrelated with the other two noise vectors. $E\left({\bf t}_{0}\right) = \overline{t_{0}}$, ${\rm cov}\left({\bf t}_{0}, {\bf t}_{0}\right) = {\bf T}_{0}$.
\end{itemize}
Above, $E$ means the expectation value (also called the first moment), $\rm{cov}$ is covariance matrix, and $\delta_{ij}^{K}$ is the Kronecker delta ($\delta_{ij}^{K} = 1$ if $i=j$, $0$ otherwise).

With this notation in hand we may now describe the system. \\ \\
\textbf{\uline{Description of system}}
\begin{equation}\label{eqn:a1}
{\bf t}_{i}=\hat{{\bf E}}_{i}{\bf t}_{i-1}+{\bf n}_{i}^{t}.
\end{equation}
The true signal at time (really, wavelength bin) $i$ is given by the true signal at bin $i-1$ evolved by the evolution matrix at bin $i$ and with the intrinsic (``process'') noise ${\bf n}_{i}^{t}$ at that time added in. Equation \ref{eqn:a1} is known as the state transition equation.

Similarly, the measurement ${\bf m}_{i}$ at bin $i$ is given by the measurement equation
\begin{equation}
{\bf m}_{i}=\hat{{\bf M}}_{i}{\bf t}_{i}+{\bf n}_{i}^{m}.
\end{equation}
We take the true signal at bin $i,$ apply the measurement matrix $\hat{{\bf M}}_{i}$, and then add in the measurement noise at that time ${\bf n}_{i}^{m}$.

Note that both of the above equations are linear functions of the state (i.e., true signal), and have additive noise. This is because in the current work we have used the linear Kalman filter; for descriptions of non--linear filters such as the Extended Kalman Filter, we offer references in \S\ref{akft}.
We now may also describe our predictions. \\ \\
\textbf{\uline{Prediction step}} \\
Our predicted signal at bin $i$ is 
\begin{equation}
{\bf t}_{i}^{{\rm pred}}=\hat{{\bf E}}_{i}{\bf t}_{i-1}^{{\rm pred}}.
\end{equation}
To predict, we simply evolve our prediction from the previous bin
using the evolution matrix. We also would like the covariance matrix
of the predicted true signal, which former we denote ${\bf T}_{i}^{{\rm pred}}.$
We have
\begin{equation}
{\bf T}_{i}^{{\rm pred}}=\hat{{\bf E}}_{i}{\bf T}_{i-1}^{{\rm upd}}{\bf \hat{E}}_{i}+{\bf N}_{i}^{t}.
\end{equation}
The right--hand side says that we take our updated covariance matrix from the $i-1^{th}$ bin, evolve it (we need two evolution matrices
since one can think of the true signal matrix as the outer product of two true signal vectors, each of which must be evolved by $\hat{{\bf E}}$), and then add the covariance matrix of the true signal noise, i.e. ${\bf N}^{t}$ is the covariance matrix of ${\bf n}^{t}.$ We have now completed our prediction step. But where does the updated signal covariance matrix ${\bf T}_{i-1}^{{\rm upd}}$ come from? How did we update this prediction from a previous one, and how will we update this prediction to move to the next one? \\ \\
\textbf{\uline{Update step}} \\
The update step is the non-trivial part of Kalman filtering: it balances
the prediction with whatever measurement we have made. The updated
signal is
\begin{equation}\label{eqn:A5}
{\bf t}_{i}^{{\rm upd}}={\bf t}_{i}^{{\rm pred}}+{\bf W}_{i}{\bf \Delta}_{i}^{m-p}.
\end{equation}
Our updated true signal (${\bf t}_{i}^{{\rm upd}}$) is the predicted one (${\bf t}_{i}^{{\rm upd}}$) plus a weight (${\bf W}_{i}$) applied to the difference between our prediction for the measured signal and
the actual signal we measured. We denote this difference as
\begin{equation}
{\bf \Delta}_{i}^{\rm m-p}\equiv{\bf m}_{i}-\hat{{\bf M}}_{i}{\bf t}_{i}^{{\rm pred}}.
\end{equation}
The superscript $m-p$ means ``measured minus predicted.'' The weight ${\bf W}_{i}$ is commonly referred to as the ``Kalman Gain''. 

In other words, our actual signal measured at bin $i$ was ${\bf m}_{i},$ but our prediction for the signal we would measure was our measurement matrix operating on our prediction ${\bf t}_{i}^{{\rm pred}}$ for the true signal. Equation \ref{eqn:A5} makes intuitive sense: to update, we begin with our predicted signal, and then correct it by adding in the weighted difference between our actually measured and predicted-to-be-measured signals (the weight can also take care of converting a measured signal to a true signal). 

The correct form of ${\bf W}_{i}$ is the only non--obvious piece of Kalman filtering. Clearly, if we had no measurement noise and if the covariance matrix for our predicted signal were the identity matrix (i.e. all bits of the signal are independent from each other and have equal error bars), then all we would need to do would be to convert ${\bf \Delta}$ from the ``measured'' space to the ``true signal'' space. In this limit, all we would need to do would be to invert our measurement matrix. We do not need anything additional to inverse--covariance--weight
since we have said that our intrinsic true signal has covariance matrix as an identity matrix; inverse--covariance weighting would thus also be an identity matrix. So we have
\[
{\bf W}_{i}^{{\rm noiseless}\; m,\;{\rm i.e.e.\; sign.} }\to\hat{{\bf M}}_{i}^{-1}.
\]
$``{\rm i.e.e.\; sign}.$'' stands for ``independent, equal error bars
signal.''

However, in practice we have measurement noise, and we also want to inverse--covariance weight the update, which is less trivial if the true signal has intrinsic covariance matrix different from the identity matrix. In this case, the predicted $measurement$ covariance matrix is
\begin{equation}\label{eqn:A7}
{\bf M}_{i}^{{\rm pred}}={\bf \hat{M}}_{i}{\bf T}_{i}^{{\rm pred}}{\bf \hat{M}}_{i}^{T};
\end{equation}
we just convert the predicted covariance matrix of the true signal into one for the measurement by applying two copies of the measurement matrix. We use two copies for the same reason as discussed earlier, regarding evolving the predicted true signal covariance matrix.

We note that the matrix ${\bf M}_{i}^{{\rm pred}}$ describes the intrinsic covariance of a noiseless measurement---it occurs because our underlying true signal has some intrinsic covariance, and we are measuring that signal. However, the full covariance of the measurement will also include a contribution from the measurement noise covariance matrix, denoted ${\bf N}_{i}^{m}$. So our total measurement covariance matrix is
\begin{equation}
{\bf M}_{i}={\bf M}_{i}^{{\rm pred}}+{\bf N}_{i}^{m}.
\end{equation}
Thus our inverse covariance weight so far looks like
\begin{equation}
{\bf W}_{i}\sim{\bf M}_{i}^{-1}=\left({\bf M}_{i}^{{\rm pred}}+{\bf N}_{i}^{m}\right)^{-1}=\left({\bf \hat{M}}_{i}{\bf T}_{i}^{{\rm pred}}{\bf \hat{M}}_{i}^{T}+{\bf N}_{i}^{m}\right)^{-1},
\end{equation}
%where we used equation (7) to obtain the second equality.
where to obtain the second equality we simply substituted equation \ref{eqn:A7} for ${\bf M}_{i}^{{\rm pred}}$. However, in the limit of no measurement noise, this weight should simply reduce to the inverse of the measurement matrix. %(Sankalp: I'm still slightly confused on this---I would think in the limit of no measurement noise, you actually still want to inverse-covariance weight! Needs a bit more thought. But derivation works if you keep going in this way.)
So we have
\begin{equation}
{\bf W}_{i}^{{\rm noiseless\;}m}=\hat{{\bf M}}_{i}^{-1}={\bf A}\left({\bf \hat{M}}_{i}{\bf T}_{i}^{{\rm pred}}{\bf \hat{M}}_{i}^{T}\right)^{-1}.
\end{equation}
The first equality comes from demanding that the weight, in the limit of no measurement noise, should just be the inverse of the measurement
matrix. We then take ${\bf A}$ to be an unknown matrix that multiplies our guess for the weight in the noiseless limit (with ${\bf N}_{i}^{m}\to0$). 

In the second equation above, ${\bf A}$ is an unknown matrix for which we wish to solve. We have a matrix equation of the form
\begin{equation}
\hat{{\bf M}}_{i}^{-1}={\bf A}{\bf B},
\end{equation}
so we apply ${\bf B}^{-1}$ from the right on each side. We have
\begin{equation}
{\bf B}^{-1}=\left({\bf \hat{M}}_{i}{\bf T}_{i}^{{\rm pred}}{\bf \hat{M}}_{i}^{T}\right),
\end{equation}
meaning
\begin{equation}
{\bf A}=\hat{{\bf M}}_{i}^{-1}\left({\bf \hat{M}}_{i}{\bf T}_{i}^{{\rm pred}}{\bf \hat{M}}_{i}^{T}\right).
\end{equation}
This simplifies to
\begin{equation}
{\bf A}={\bf T}_{i}^{{\rm pred}}\hat{{\bf M}}_{i}^{T}.
\end{equation}
So our full weight is now
\begin{equation}\label{eqn:A15}
{\bf W}_{i}={\bf T}_{i}^{{\rm pred}}\hat{{\bf M}}_{i}^{T}\left({\bf \hat{M}}_{i}{\bf T}_{i}^{{\rm pred}}{\bf \hat{M}}_{i}^{T}\right)^{-1}.
\end{equation}
We now insert the weight from equation \ref{eqn:A15} into equation \ref{eqn:A5} to update our prediction.

\section{Tables}\label{nine_tables}
Incomplete tables are displayed here. Complete tables are available online.\footnote{\url{https://github.com/astrogilda/stellar_denoising}}.

\begin{table*}%{\textwidth}%[htbp]
\centering
\caption{The means and standard deviations for the error metrics for different denoising methods applied to the stellar absorption spectrum of \emph{Star 1} corrupted with Poisson noise to get a PSNR of $5.0$ (the results for the other two stars show similar trends). The complete table, along with a detailed caption, is available online. See Figs. \ref{fig:fig4}--\ref{fig:fig7} (a) at the {\it x--axis} value of 5 for a graphical illustration of this table.}
\label{table:tab11}
\begin{tabularx}{\textwidth}{XCCCC}
\toprule
\parnoteclear % tabularx will otherwise add each note thrice
                                     & {$\bm{L^{1}}$}                    & {$\bm{L^{2}}$}                    & {$\bm{L^{\infty}}$}                  & {$\bm{SSIM}$}                  \\ \midrule
\multicolumn{1}{l}{\textbf{Noisy data}}     & \multicolumn{1}{c}{$0.113 \pm 1.5\%$} & \multicolumn{1}{c}{$0.144 \pm 1.3\%$} & \multicolumn{1}{c}{$0.595 \pm 10.2\%$} & \multicolumn{1}{c}{$0.385 \pm 1.8\%$} \\ \midrule
\multicolumn{1}{l}{\textbf{Moving Mean}\parnote{w = $(7, 7, 8, 7)$}}      & \multicolumn{1}{c}{$0.048 \pm 2.8\%$} & \multicolumn{1}{c}{$0.060 \pm 2.1\%$} & \multicolumn{1}{c}{$0.246 \pm 11.9\%$} & \multicolumn{1}{c}{$0.834 \pm 0.8\%$} \\ \midrule
\multicolumn{1}{l}{\textbf{Exp. Smoothing}\parnote{span = $(4, 4, 5, 6)$}} & \multicolumn{1}{c}{$\textit{0.067} \pm \textit{1.6\%}$} & \multicolumn{1}{c}{$\textit{0.086} \pm \textit{1.7\%}$} & \multicolumn{1}{c}{$\textit{0.328} \pm \textit{5.6\%}$} & \multicolumn{1}{c}{$\textit{0.690} \pm \textit{1.1\%}$} \\ \midrule
\multicolumn{1}{l}{\textbf{FFT, Gaussian kernel}\parnote{$\sigma$ = $(2, 2, 2, 3)$}}     & \multicolumn{1}{c}{$\textbf{0.047} \pm \textbf{2.1\%}$} & \multicolumn{1}{c}{$\textbf{0.059} \pm \textbf{2.5\%}$} & \multicolumn{1}{c}{$\textbf{0.219} \pm \textbf{5.1\%}$} & \multicolumn{1}{c}{$\textbf{0.853} \pm \textbf{0.6\%}$} \\ \midrule
\multicolumn{1}{l}{\textbf{Wavelet + AKFThresh v1}}  & \multicolumn{1}{c}{$0.050 \pm 1.7\%$} & \multicolumn{1}{c}{$0.064 \pm 1.7\%$} & \multicolumn{1}{c}{$0.256 \pm 8.6\%$} & \multicolumn{1}{c}{$0.817 \pm 0.6\%$} \\ \bottomrule
\end{tabularx}
\vspace{-2ex}
\parnotes

\end{table*}

%\begin{subtable}{\textwidth}
%\centering
%\begin{tabular}{@{}lcccc@{}}
\begin{table*}
\centering
\caption{Same as Table \ref{table:tab11}, but using the tenth--best parameters for each denoising method. See Figs. \ref{fig:fig4}--\ref{fig:fig7} (b) at the {\it x--axis} value of 5 for a graphical illustration of this table.}
\label{table:tab12}
\begin{tabularx}{\textwidth}{XCCCC}
\toprule
\parnoteclear % tabularx will otherwise add each note thrice
                                     & {$\bm{L^{1}}$}                    & {$\bm{L^{2}}$}                    & {$\bm{L^{\infty}}$}                  & {$\bm{SSIM}$}                  \\ \midrule
\multicolumn{1}{l}{\textbf{Noisy data}}     & \multicolumn{1}{c}{$0.113 \pm 1.1\%$} & \multicolumn{1}{c}{$0.145 \pm 1.1\%$} & \multicolumn{1}{c}{$0.655 \pm 14.2\%$} & \multicolumn{1}{c}{$0.382 \pm 1.1\%$} \\ \midrule
\multicolumn{1}{l}{\textbf{Moving Mean}\parnote{w = $(13, 8, 12, 4)$ }}                                      & \multicolumn{1}{c}{$0.060 \pm 1.2\%$} & \multicolumn{1}{c}{$0.076 \pm 1.1\%$} & \multicolumn{1}{c}{$0.318 \pm 3.6\%$} & \multicolumn{1}{c}{$0.726 \pm 1.1\%$} \\ \midrule
\multicolumn{1}{l}{\textbf{Exp. Smoothing}\parnote{span = $(11, 11, 12, 13)$}}             & \multicolumn{1}{c}{$0.087 \pm 1.4\%$} & \multicolumn{1}{c}{$0.108 \pm 1.2\%$} & \multicolumn{1}{c}{$0.380 \pm 3.4\%$} & \multicolumn{1}{c}{$0.654 \pm 0.5\%$} \\ \midrule
\multicolumn{1}{l}{\textbf{FFT, Gaussian kernel}\parnote{$\sigma = (11, 11, 11, 11)$}}              & \multicolumn{1}{c}{$0.098 \pm 0.6\%$} & \multicolumn{1}{c}{$0.123 \pm 0.4\%$} & \multicolumn{1}{c}{$\textit{0.411} \pm \textit{2.3\%}$} & \multicolumn{1}{c}{$\textit{0.614} \pm \textit{0.1\%}$} \\ \midrule
\multicolumn{1}{l}{\textbf{Wavelet + AKFThresh v1}} & \multicolumn{1}{c}{$0.049 \pm 2.4\%$} & \multicolumn{1}{c}{$0.063 \pm 2.9\%$} & \multicolumn{1}{c}{$0.253 \pm 6.5\%$} & \multicolumn{1}{c}{$0.815 \pm 1.2\%$} \\ \bottomrule
\end{tabularx}
%\caption{}%{PSNR = 1.0}
%\label{table:tab12} 
\vspace{-2ex}
\parnotes
%\end{subtable}
\end{table*}

%\bigskip
\begin{table*}
%\begin{subtable}{\textwidth}
\centering
\caption{Same as Table \ref{table:tab12}, but using the twentieth--best parameters for the error metrics. See Figs. \ref{fig:fig4}--\ref{fig:fig7} (c) at the {\it x--axis} value of 5 for a graphical illustration of this table.}
\label{table:tab13}
\begin{tabularx}{\textwidth}{XCCCC}
\toprule
\parnoteclear % tabularx will otherwise add each note thrice
                                     & {$\bm{L^{1}}$}                    & {$\bm{L^{2}}$}                    & {$\bm{L^{\infty}}$}                  & {$\bm{SSIM}$}                  \\ \midrule
\multicolumn{1}{l}{\textbf{Noisy data}}     & \multicolumn{1}{c}{$0.112 \pm 0.8\%$} & \multicolumn{1}{c}{$0.145 \pm 0.9\%$} & \multicolumn{1}{c}{$0.636 \pm 8.0\%$} & \multicolumn{1}{c}{$0.384 \pm 0.9\%$} \\ \midrule
\multicolumn{1}{l}{\textbf{Moving Mean}\parnote{w = ($21, 21, 22, 22$)}}                                      & \multicolumn{1}{c}{$0.084 \pm 0.8\%$} & \multicolumn{1}{c}{$0.107 \pm 0.6\%$} & \multicolumn{1}{c}{$0.392 \pm 4.7\%$} & \multicolumn{1}{c}{$0.597 \pm 0.6\%$} \\ \midrule
\multicolumn{1}{l}{\textbf{Exp. Smoothing}\parnote{span = ($21, 21, 21, 22$)}}            & \multicolumn{1}{c}{$0.108 \pm 0.7\%$} & \multicolumn{1}{c}{$0.135 \pm 0.5\%$} & \multicolumn{1}{c}{$0.413 \pm 4.1\%$} & \multicolumn{1}{c}{$0.616 \pm 0.7\%$} \\ \midrule
\multicolumn{1}{l}{\textbf{FFT, Gaussian kernel}\parnote{$\sigma = (21, 21, 21, 21)$}}             & \multicolumn{1}{c}{$\textit{0.122} \pm \textit{0.4\%}$} & \multicolumn{1}{c}{$\textit{0.150} \pm \textit{0.3\%}$} & \multicolumn{1}{c}{$\textit{0.478} \pm \textit{1.8\%}$} & \multicolumn{1}{c}{$\textit{0.567} \pm \textit{0.1\%}$} \\ \midrule
\multicolumn{1}{l}{\textbf{Wavelet + AKFThresh v1}}  & \multicolumn{1}{c}{$0.050 \pm 2.7\%$} & \multicolumn{1}{c}{$0.064 \pm 2.4\%$} & \multicolumn{1}{c}{$0.267 \pm 11.0\%$} & \multicolumn{1}{c}{$0.814 \pm 0.8\%$} \\ \bottomrule
\end{tabularx}
%\caption{}%{PSNR = 1.0}
%\label{table:tab13} 
\vspace{-2ex}
\parnotes
%\end{subtable}
%\end{minipage}
\end{table*}

%\bigskip
\begin{table*}
%\begin{subtable}{\textwidth}
\centering
\caption{Same as Table \ref{table:tab11}, but with PSNR of $10$. See Figs. \ref{fig:fig4}--\ref{fig:fig7} (a) at the {\it x--axis} value of 10 for a graphical illustration of this table.}
\label{table:tab21}
\begin{tabularx}{\textwidth}{XCCCC}
\toprule
\parnoteclear % tabularx will otherwise add each note thrice
                                     & {$\bm{L^{1}}$}                    & {$\bm{L^{2}}$}                    & {$\bm{L^{\infty}}$}                  & {$\bm{SSIM}$}                  \\ \midrule
\multicolumn{1}{l}{\textbf{Noisy data}}     & \multicolumn{1}{c}{$0.056 \pm 1.4\%$} & \multicolumn{1}{c}{$0.072 \pm 1.3\%$} & \multicolumn{1}{c}{$0.324 \pm 4.3\%$} & \multicolumn{1}{c}{$0.662 \pm 1.0\%$} \\ \midrule
\multicolumn{1}{l}{\textbf{Moving Mean}\parnote{w = ($5, 5, 4, 7$)}}                                      & \multicolumn{1}{c}{$0.028 \pm 1.6\%$} & \multicolumn{1}{c}{$0.035 \pm 1.7\%$} & \multicolumn{1}{c}{$0.156 \pm 7.9\%$} & \multicolumn{1}{c}{$0.922 \pm 0.3\%$} \\ \midrule
\multicolumn{1}{l}{\textbf{Exp. Smoothing}\parnote{span = ($3, 3, 3, 4$)}}            & \multicolumn{1}{c}{$\textit{0.042} \pm \textit{1.6\%}$} & \multicolumn{1}{c}{$\textit{0.054} \pm \textit{1.8\%}$} & \multicolumn{1}{c}{$\textit{0.225} \pm \textit{12.4\%}$} & \multicolumn{1}{c}{$\textit{0.827} \pm \textit{0.6\%}$} \\ \midrule
\multicolumn{1}{l}{\textbf{FFT, Gaussian kernel}\parnote{$\sigma = (2, 2, 2, 2)$}}             & \multicolumn{1}{c}{$0.028 \pm 2.6\%$} & \multicolumn{1}{c}{$0.036 \pm 2.9\%$} & \multicolumn{1}{c}{$0.152 \pm 22.9\%$} & \multicolumn{1}{c}{$0.932 \pm 0.3\%$} \\ \midrule
\multicolumn{1}{l}{\textbf{Wavelet + AKFThresh v1}}  & \multicolumn{1}{c}{$0.028 \pm 2.1\%$} & \multicolumn{1}{c}{$0.037 \pm 2.2\%$} & \multicolumn{1}{c}{$0.159 \pm 11.1\%$} & \multicolumn{1}{c}{$0.911 \pm 0.3\%$} \\ \bottomrule
\end{tabularx}
%\caption{}%{PSNR = 1.0}
%\label{table:tab13} 
\vspace{-2ex}
\parnotes
%\end{subtable}
%\end{minipage}
\end{table*}

\begin{table*}
%\begin{subtable}{\textwidth}
\centering
\caption{Same as Table \ref{table:tab12}, but with PSNR of $10$. See Figs. \ref{fig:fig4}--\ref{fig:fig7} (b) at the {\it x--axis} value of 10 for a graphical illustration of this table.}
\label{table:tab22}
\begin{tabularx}{\textwidth}{XCCCC}
\toprule
\parnoteclear % tabularx will otherwise add each note thrice
                                     & {$\bm{L^{1}}$}                    & {$\bm{L^{2}}$}                    & {$\bm{L^{\infty}}$}                  & {$\bm{SSIM}$}                  \\ \midrule
\multicolumn{1}{l}{\textbf{Noisy data}}     & \multicolumn{1}{c}{$0.056 \pm 1.3\%$} & \multicolumn{1}{c}{$0.073 \pm 1.1\%$} & \multicolumn{1}{c}{$0.330 \pm 7.6\%$} & \multicolumn{1}{c}{$0.661 \pm 0.7\%$} \\ \midrule
\multicolumn{1}{l}{\textbf{Moving Mean}\parnote{w = ($11, 11, 11, 12$)}}                                      & \multicolumn{1}{c}{$0.044 \pm 1.0\%$} & \multicolumn{1}{c}{$0.056 \pm 0.8\%$} & \multicolumn{1}{c}{$0.235 \pm 7.8\%$} & \multicolumn{1}{c}{$0.799 \pm 0.2\%$} \\ \midrule
\multicolumn{1}{l}{\textbf{Exp. Smoothing}\parnote{span = ($11, 11, 11, 12$)}}            & \multicolumn{1}{c}{$0.081 \pm 0.5\%$} & \multicolumn{1}{c}{$0.102 \pm 0.5\%$} & \multicolumn{1}{c}{$0.328 \pm 3.4\%$} & \multicolumn{1}{c}{$0.713 \pm 0.3\%$} \\ \midrule
\multicolumn{1}{l}{\textbf{FFT, Gaussian kernel}\parnote{$\sigma = (11, 11, 11, 11)$}}             & \multicolumn{1}{c}{$\textit{0.096} \pm \textit{0.4\%}$} & \multicolumn{1}{c}{$\textit{0.121} \pm \textit{0.2\%}$} & \multicolumn{1}{c}{$\textit{0.395} \pm \textit{1.5\%}$} & \multicolumn{1}{c}{$\textit{0.616} \pm \textit{0.1\%}$} \\ \midrule
\multicolumn{1}{l}{\textbf{Wavelet + AKFThresh v1}}  & \multicolumn{1}{c}{$0.028 \pm 2.0\%$} & \multicolumn{1}{c}{$0.036 \pm 1.9\%$} & \multicolumn{1}{c}{$0.157 \pm 9.4\%$} & \multicolumn{1}{c}{$0.911 \pm 0.3\%$} \\ \bottomrule
\end{tabularx}
%\caption{}%{PSNR = 1.0}
%\label{table:tab13} 
\vspace{-2ex}
\parnotes
%\end{subtable}
%\end{minipage}
\end{table*}

\begin{table*}
%\begin{subtable}{\textwidth}
\centering
\caption{Same as Table \ref{table:tab13}, but with PSNR of $10$. See Figs. \ref{fig:fig4}--\ref{fig:fig7} (c) at the {\it x--axis} value of 10 for a graphical illustration of this table.}
\label{table:tab23}
\begin{tabularx}{\textwidth}{XCCCC}
\toprule
\parnoteclear % tabularx will otherwise add each note thrice
                                     & {$\bm{L^{1}}$}                    & {$\bm{L^{2}}$}                    & {$\bm{L^{\infty}}$}                  & {$\bm{SSIM}$}                  \\ \midrule
\multicolumn{1}{l}{\textbf{Noisy data}}     & \multicolumn{1}{c}{$0.056 \pm 0.8\%$} & \multicolumn{1}{c}{$0.072 \pm 0.9\%$} & \multicolumn{1}{c}{$0.313 \pm 8.7\%$} & \multicolumn{1}{c}{$0.664 \pm 0.7\%$} \\ \midrule
\multicolumn{1}{l}{\textbf{Moving Mean}\parnote{w = ($21, 21, 21, 21$)}}                                      & \multicolumn{1}{c}{$0.081 \pm 0.4\%$} & \multicolumn{1}{c}{$0.104 \pm 0.2\%$} & \multicolumn{1}{c}{$0.370 \pm 5.7\%$} & \multicolumn{1}{c}{$0.614 \pm 0.2\%$} \\ \midrule
\multicolumn{1}{l}{\textbf{Exp. Smoothing}\parnote{span = ($21, 21, 21, 21$)}}            & \multicolumn{1}{c}{$0.107 \pm 0.5\%$} & \multicolumn{1}{c}{$0.131 \pm 0.4\%$} & \multicolumn{1}{c}{$0.398 \pm 2.4\%$} & \multicolumn{1}{c}{$0.632 \pm 0.4\%$} \\ \midrule
\multicolumn{1}{l}{\textbf{FFT, Gaussian kernel}\parnote{$\sigma = (21, 21, 21, 21)$}}             & \multicolumn{1}{c}{$\textit{0.121} \pm \textit{0.1\%}$} & \multicolumn{1}{c}{$\textit{0.149} \pm \textit{0.1\%}$} & \multicolumn{1}{c}{$\textit{0.475} \pm \textit{1.3\%}$} & \multicolumn{1}{c}{$\textit{0.568} \pm \textit{0.0\%}$} \\ \midrule
\multicolumn{1}{l}{\textbf{Wavelet + AKFThresh v1}}  & \multicolumn{1}{c}{$0.028 \pm 1.5\%$} & \multicolumn{1}{c}{$0.036 \pm 1.9\%$} & \multicolumn{1}{c}{$0.154 \pm 7.3\%$} & \multicolumn{1}{c}{$0.913 \pm 0.3\%$} \\ \bottomrule
\end{tabularx}
%\caption{}%{PSNR = 1.0}
%\label{table:tab13} 
\vspace{-2ex}
\parnotes
%\end{subtable}
%\end{minipage}
\end{table*}

\begin{table*}
%\begin{subtable}{\textwidth}
\centering
\caption{Same as Tables \ref{table:tab11} and \ref{table:tab21}, but with PSNR of $20$. See Figs. \ref{fig:fig4}--\ref{fig:fig7} (a) at the {\it x--axis} value of 20 for a graphical illustration of this table.}
\label{table:tab31}
\begin{tabularx}{\textwidth}{XCCCC}
\toprule
\parnoteclear % tabularx will otherwise add each note thrice
                                     & {$\bm{L^{1}}$}                    & {$\bm{L^{2}}$}                    & {$\bm{L^{\infty}}$}                  & {$\bm{SSIM}$}                  \\ \midrule
\multicolumn{1}{l}{\textbf{Noisy data}}     & \multicolumn{1}{c}{$0.028 \pm 1.2\%$} & \multicolumn{1}{c}{$0.036 \pm 1.2\%$} & \multicolumn{1}{c}{$0.151 \pm 10.8\%$} & \multicolumn{1}{c}{$0.869 \pm 0.3\%$} \\ \midrule
\multicolumn{1}{l}{\textbf{Moving Mean}\parnote{w = ($3, 5, 3, 5$)}}                                      & \multicolumn{1}{c}{$0.017 \pm 1.2\%$} & \multicolumn{1}{c}{$0.021 \pm 1.2\%$} & \multicolumn{1}{c}{$0.089 \pm 4.2\%$} & \multicolumn{1}{c}{$0.969 \pm 0.1\%$} \\ \midrule
\multicolumn{1}{l}{\textbf{Exp. Smoothing}\parnote{span = ($2, 2, 2, 2$)}}            & \multicolumn{1}{c}{$\textit{0.024} \pm \textit{1.0\%}$} & \multicolumn{1}{c}{$\textit{0.031} \pm \textit{0.9\%}$} & \multicolumn{1}{c}{$0.124 \pm 8.7\%$} & \multicolumn{1}{c}{$\textit{0.922} \pm \textit{0.2\%}$} \\ \midrule
\multicolumn{1}{l}{\textbf{FFT, Gaussian kernel}\parnote{$\sigma = (2, 2, 2, 2)$}}             & \multicolumn{1}{c}{$0.021 \pm 1.2\%$} & \multicolumn{1}{c}{$0.027 \pm 1.2\%$} & \multicolumn{1}{c}{$\textit{0.132} \pm \textit{14.5\%}$} & \multicolumn{1}{c}{$0.958 \pm 0.1\%$} \\ \midrule
\multicolumn{1}{l}{\textbf{Wavelet + AKFThresh v1}}  & \multicolumn{1}{c}{$0.016 \pm 2.4\%$} & \multicolumn{1}{c}{$0.020 \pm 2.1\%$} & \multicolumn{1}{c}{$0.083 \pm 6.9\%$} & \multicolumn{1}{c}{$0.966 \pm 0.1\%$} \\ \bottomrule
\end{tabularx}
%\caption{}%{PSNR = 1.0}
%\label{table:tab13} 
\vspace{-2ex}
\parnotes
%\end{subtable}
%\end{minipage}
\end{table*}

\begin{table*}
%\begin{subtable}{\textwidth}
\centering
\caption{Same as Tables \ref{table:tab12} and \ref{table:tab22}, but with PSNR of $20$. See Figs. \ref{fig:fig4}--\ref{fig:fig7} (b) at the {\it x--axis} value of 20 for a graphical illustration of this table.}
\label{table:tab32}
\begin{tabularx}{\textwidth}{XCCCC}
\toprule
\parnoteclear % tabularx will otherwise add each note thrice
                                     & {$\bm{L^{1}}$}                    & {$\bm{L^{2}}$}                    & {$\bm{L^{\infty}}$}                  & {$\bm{SSIM}$}                  \\ \midrule
\multicolumn{1}{l}{\textbf{Noisy data}}     & \multicolumn{1}{c}{$0.028 \pm 1.1\%$} & \multicolumn{1}{c}{$0.036 \pm 1.0\%$} & \multicolumn{1}{c}{$0.152 \pm 7.9\%$} & \multicolumn{1}{c}{$0.868 \pm 0.4\%$} \\ \midrule
\multicolumn{1}{l}{\textbf{Moving Mean}\parnote{w = ($11, 11, 11, 11$)}}                                      & \multicolumn{1}{c}{$0.041 \pm 0.6\%$} & \multicolumn{1}{c}{$0.053 \pm 0.4\%$} & \multicolumn{1}{c}{$0.237 \pm 5.1\%$} & \multicolumn{1}{c}{$0.842 \pm 0.1\%$} \\ \midrule
\multicolumn{1}{l}{\textbf{Exp. Smoothing}\parnote{span = ($11, 11, 11, 11$)}}            & \multicolumn{1}{c}{$0.080 \pm 0.5\%$} & \multicolumn{1}{c}{$0.100 \pm 0.4\%$} & \multicolumn{1}{c}{$0.310 \pm 1.9\%$} & \multicolumn{1}{c}{$0.727 \pm 0.2\%$} \\ \midrule
\multicolumn{1}{l}{\textbf{FFT, Gaussian kernel}\parnote{$\sigma = (11, 11, 11, 11)$}}             & \multicolumn{1}{c}{$\textit{0.096} \pm \textit{0.2\%}$} & \multicolumn{1}{c}{$\textit{0.121} \pm \textit{0.1\%}$} & \multicolumn{1}{c}{$\textit{0.397} \pm \textit{1.0\%}$} & \multicolumn{1}{c}{$\textit{0.616} \pm \textit{0.0\%}$} \\ \midrule
\multicolumn{1}{l}{\textbf{Wavelet + AKFThresh v1}}  & \multicolumn{1}{c}{$0.016 \pm 1.4\%$} & \multicolumn{1}{c}{$0.020 \pm 1.6\%$} & \multicolumn{1}{c}{$0.091 \pm 9.0\%$} & \multicolumn{1}{c}{$0.965 \pm 0.1\%$} \\ \bottomrule
\end{tabularx}
%\caption{}%{PSNR = 1.0}
%\label{table:tab13} 
\vspace{-2ex}
\parnotes
%\end{subtable}
%\end{minipage}
\end{table*}

\begin{table*}
%\begin{subtable}{\textwidth}
\centering
\caption{Same as Tables \ref{table:tab13} and \ref{table:tab23}, but with PSNR of noisy signal $20$. See Figs. \ref{fig:fig4}--\ref{fig:fig7} (c) at the {\it x--axis} value of 20 for a graphical illustration of this table.}
\label{table:tab33}
\begin{tabularx}{\textwidth}{XCCCC}
\toprule
\parnoteclear % tabularx will otherwise add each note thrice
                                     & {$\bm{L^{1}}$}                    & {$\bm{L^{2}}$}                    & {$\bm{L^{\infty}}$}                  & {$\bm{SSIM}$}                  \\ \midrule
\multicolumn{1}{l}{\textbf{Noisy data}}     & \multicolumn{1}{c}{$0.028 \pm 1.0\%$} & \multicolumn{1}{c}{$0.001 \pm 0.9\%$} & \multicolumn{1}{c}{$0.165 \pm 6.7\%$} & \multicolumn{1}{c}{$0.868 \pm 0.2\%$} \\ \midrule
\multicolumn{1}{l}{\textbf{Moving Mean}\parnote{w = ($21, 21, 21, 21$)}}                                      & \multicolumn{1}{c}{$0.080 \pm 0.2\%$} & \multicolumn{1}{c}{$0.103 \pm 0.1\%$} & \multicolumn{1}{c}{$0.354 \pm 2.8\%$} & \multicolumn{1}{c}{$0.615 \pm 0.1\%$} \\ \midrule
\multicolumn{1}{l}{\textbf{Exp. Smoothing}\parnote{span = ($21, 21, 21, 21$)}}            & \multicolumn{1}{c}{$0.106 \pm 0.3\%$} & \multicolumn{1}{c}{$0.131 \pm 0.2\%$} & \multicolumn{1}{c}{$0.380 \pm 0.9\%$} & \multicolumn{1}{c}{$0.637 \pm 0.2\%$} \\ \midrule
\multicolumn{1}{l}{\textbf{FFT, Gaussian kernel}\parnote{$\sigma = (21, 21, 21, 21)$}}             & \multicolumn{1}{c}{$\textit{0.121} \pm \textit{0.1\%}$} & \multicolumn{1}{c}{$\textit{0.149} \pm \textit{0.0\%}$} & \multicolumn{1}{c}{$\textit{0.476} \pm \textit{0.8\%}$} & \multicolumn{1}{c}{$\textit{0.568} \pm \textit{0.0\%}$} \\ \midrule
\multicolumn{1}{l}{\textbf{Wavelet + AKFThresh v1}}  & \multicolumn{1}{c}{$0.016 \pm 1.6\%$} & \multicolumn{1}{c}{$\textbf{0.020} \pm \textbf{1.5\%}$} & \multicolumn{1}{c}{$\textbf{0.082} \pm \textbf{7.2\%}$} & \multicolumn{1}{c}{$0.967 \pm 0.2\%$} \\ \bottomrule
\end{tabularx}
%\caption{}%{PSNR = 1.0}
%\label{table:tab13} 
\vspace{-2ex}
\parnotes
%\end{subtable}
%\end{minipage}
\end{table*}

%Tables are displayed on the following pages.

\section{Denoising Comparison}
\label{appendix_figs}

We present the comparative denoising results of our proposed algorithms with state--of--the--art algorithms, for peak signal--to--noise ratios of 10.0 and 20.0. See also Fig. \ref{fig:fig10} for comparison.

\begin{figure*}
\centering
%\begin{center}
\includegraphics[width=.95\linewidth]{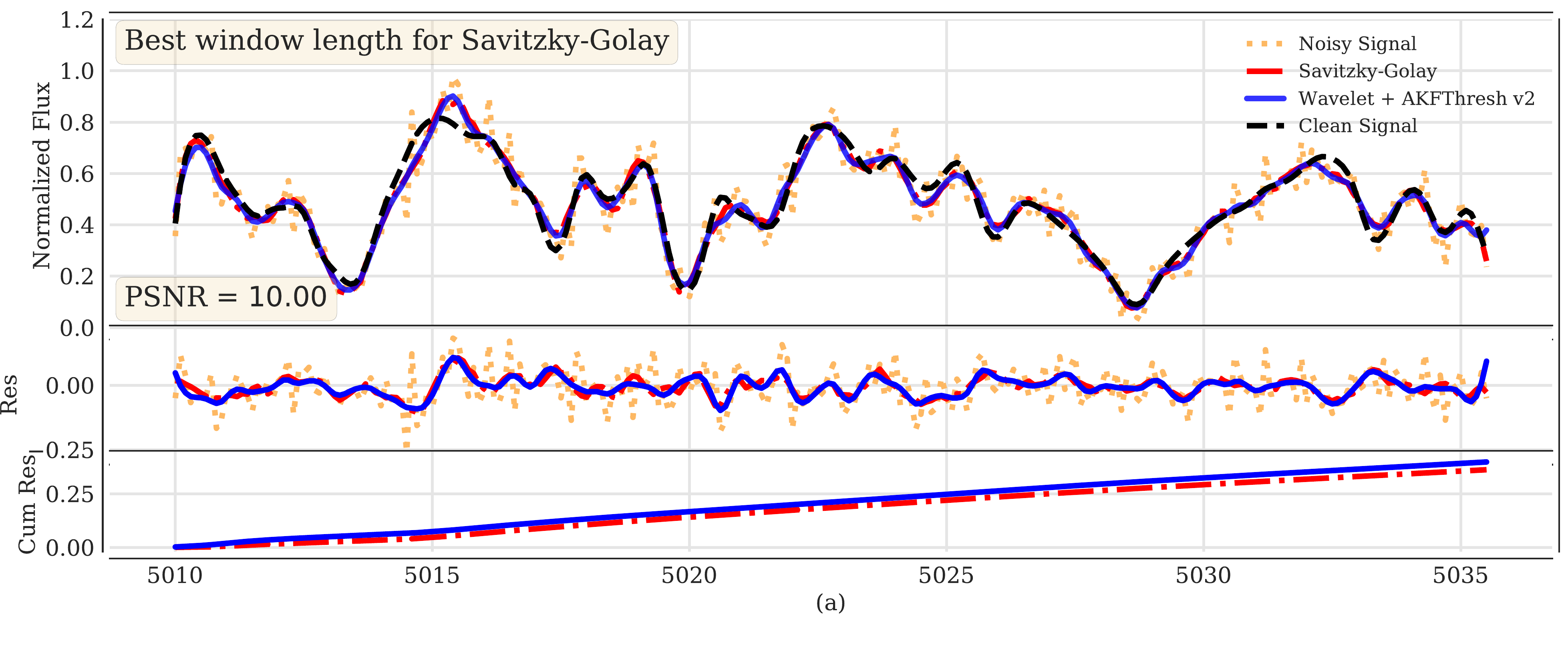}
\includegraphics[width=.95\linewidth]{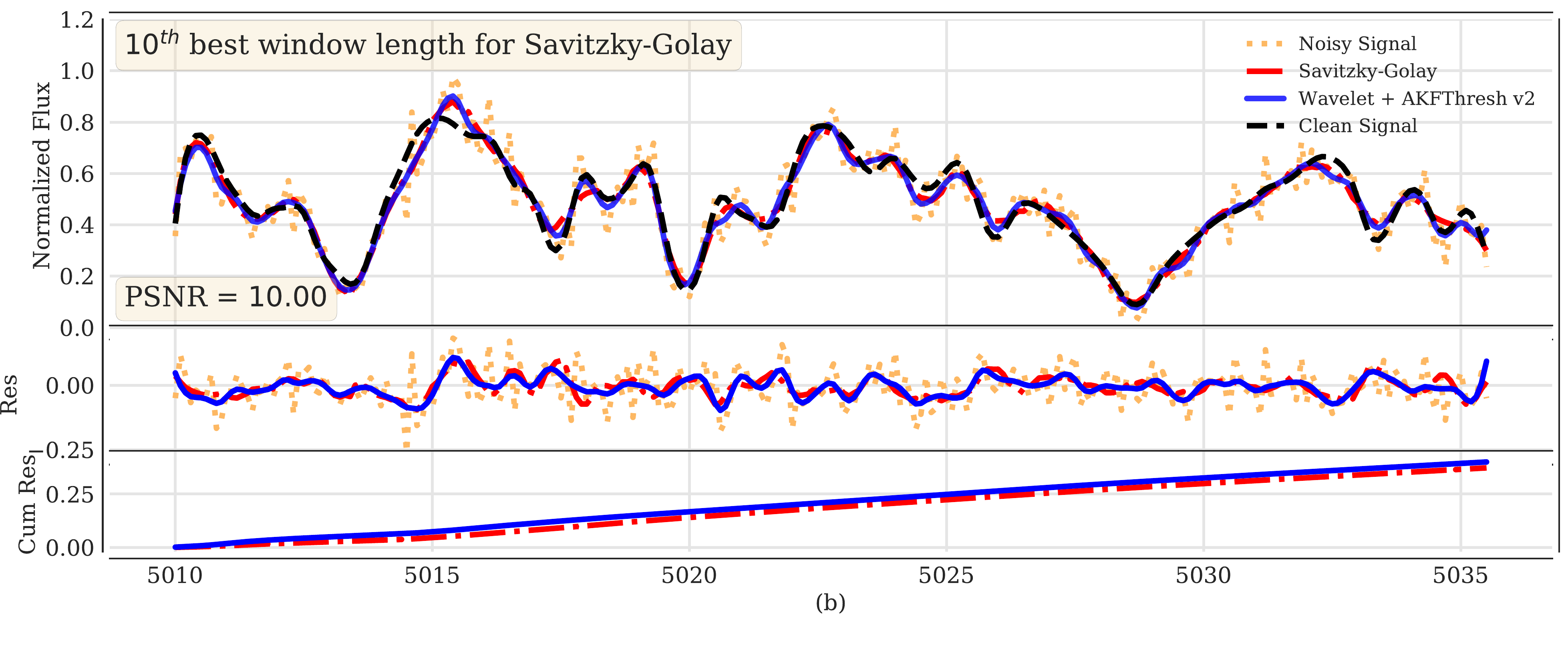}
\includegraphics[width=.95\linewidth]{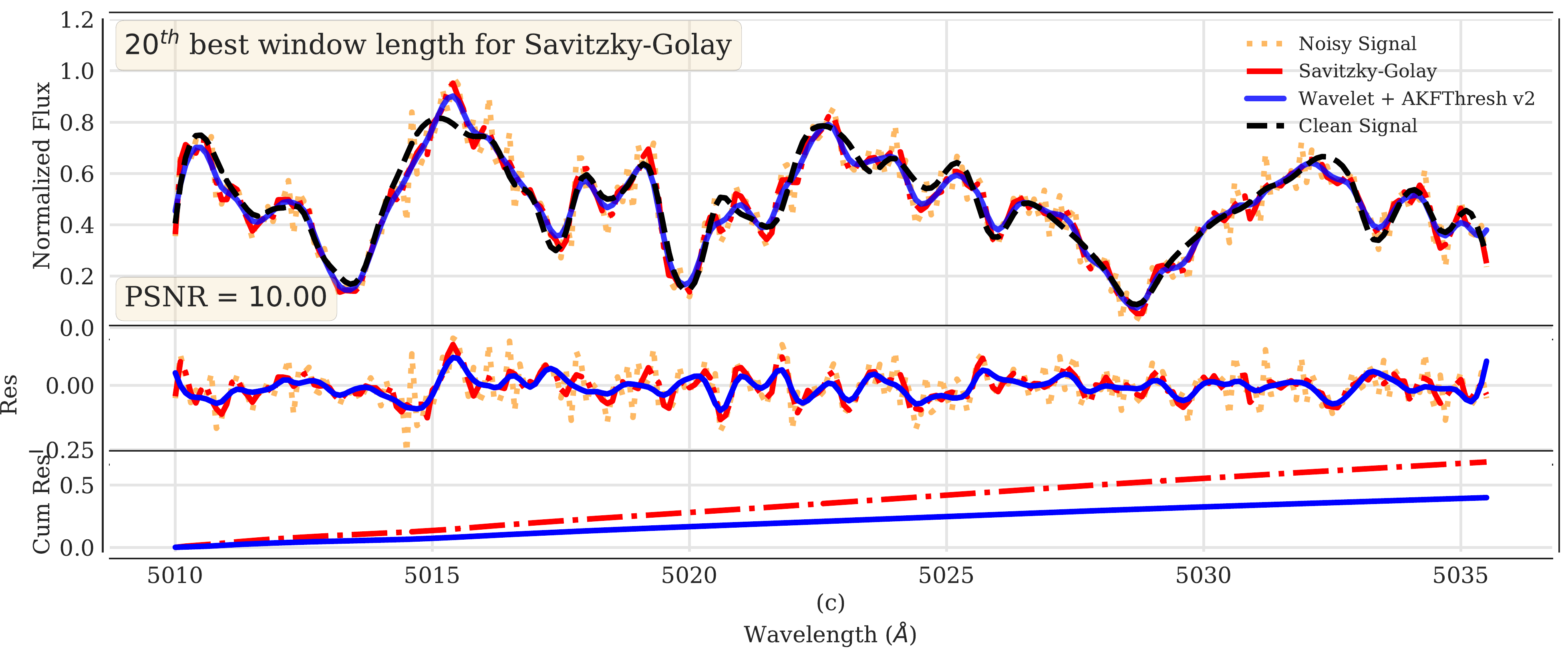}
\caption{Same as Fig. \ref{fig:fig10}, but with PSNR $= 10.0$. Here our method and SG are more comparable when SG has its best and tenth-best window lengths, but with the twentieth-best window length our method clearly outperforms SG, as shown by the lower cumulative residual (lowermost panel of sub-figure {\it (c)}). Interestingly, higher PSNR seems to make SG and our method more similar for tenth and twentieth-best window choices, but separates the two approaches more (to the advantage of ours) for twentieth-best window size relative to the same plots for PSNR = $5.0$.}
\label{fig:fig11}
%\end{center}
\end{figure*}

\begin{figure*}
\centering
%\begin{center}
\includegraphics[width=.95\linewidth]{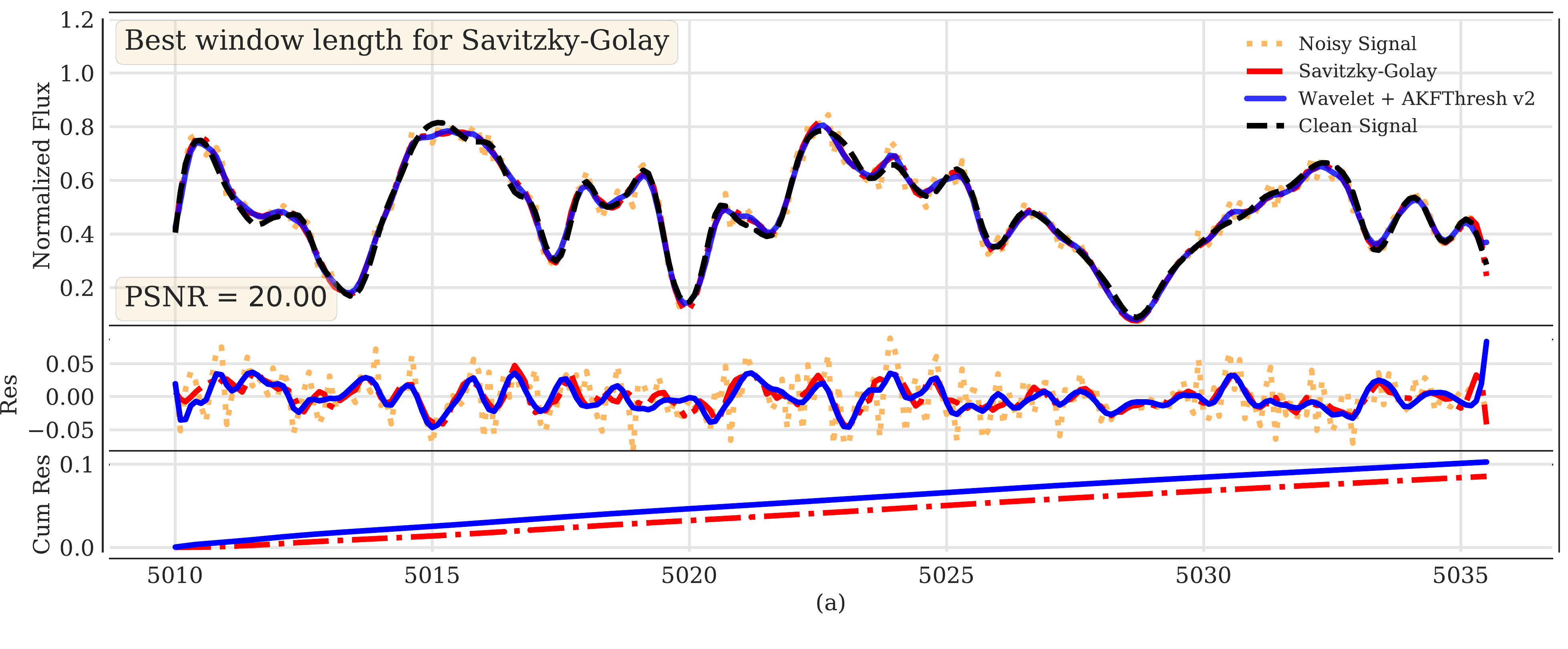}
\includegraphics[width=.95\linewidth]{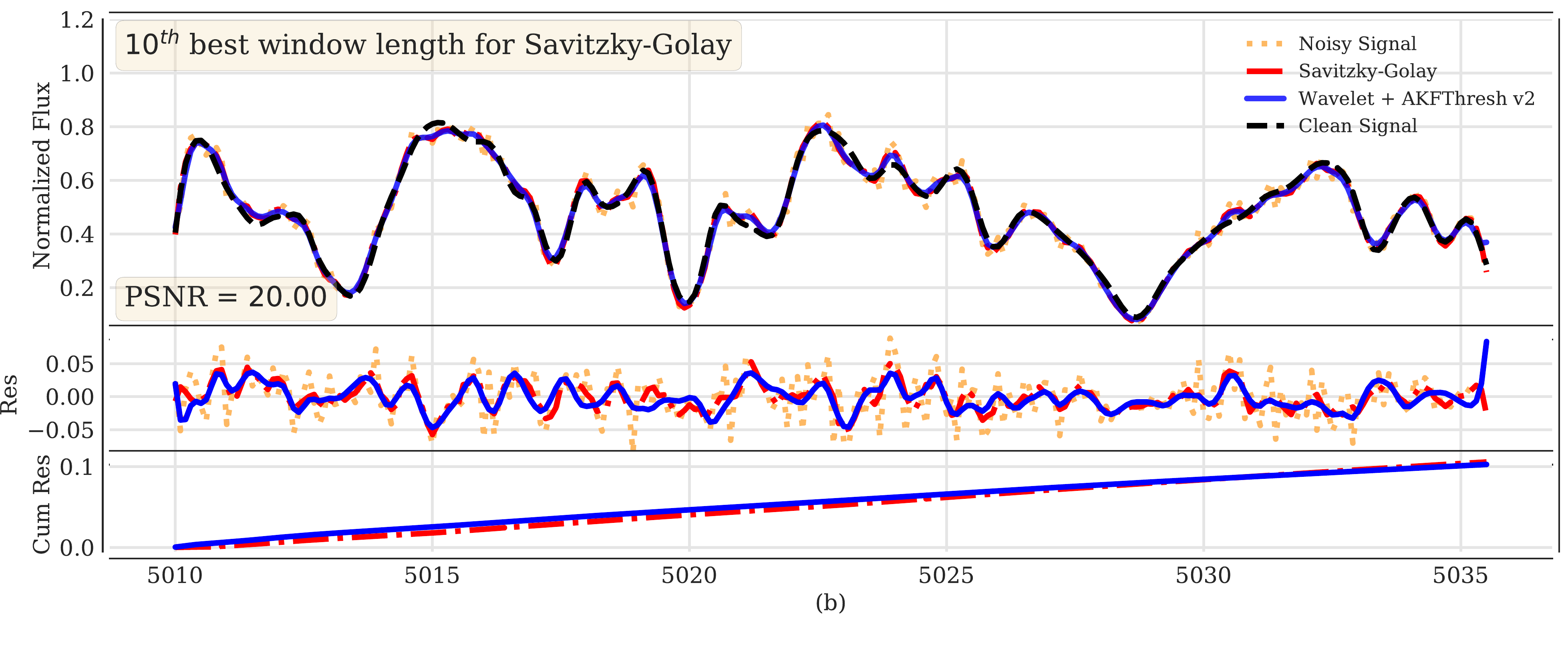}
\includegraphics[width=.95\linewidth]{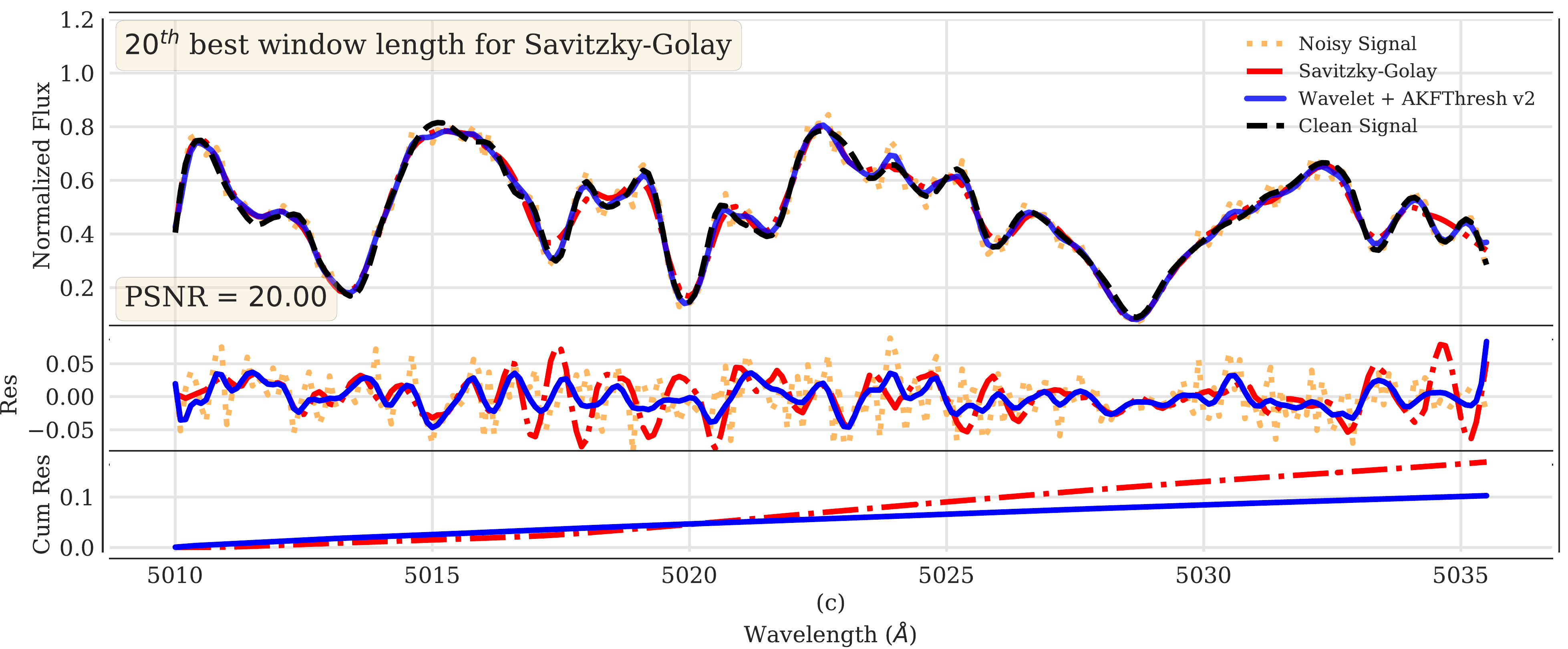}
\caption{Same as Fig. \ref{fig:fig10}, but with PSNR $= 20.0$. Here SG actually slightly outperforms our method when its best window length is used; the methods are equal at SG's tenth-best window length, and our method performs noticeably better when SG has its twentieth-best window length (lowermost panel in sub-figure {\it (c)}). Overall, comparing Figs. \ref{fig:fig10}, \ref{fig:fig11}, and this one, our method's outperformance vs. SG with twentieth-best window length is similar for PSNR of $10.0$ and $20.0$ and slightly smaller for PSNR of $5.0$.}
\label{fig:fig12}
%\end{center}
\end{figure*}

% Don't change these lines
%\bsp	% typesetting comment
\label{lastpage}
\end{document}